\documentclass{aa}  
\usepackage{graphicx}
\usepackage{subfigure}
\usepackage{txfonts}
\begin{document} 

   \title{Subarcsecond international LOFAR radio images of the M82 nucleus at 118\,MHz and 154\,MHz}
      \author{E. Varenius \inst{\ref{inst:chalmers}}
          \and 
          J. E. Conway \inst{\ref{inst:chalmers}}
          \and 
          I. Martí-Vidal \inst{\ref{inst:chalmers}}
          \and 
          R. Beswick \inst{\ref{inst:UK}}
          \and
          A. T. Deller \inst{\ref{inst:ASTRON}}
          \and
          O. Wucknitz \inst{\ref{inst:mpifr}}
          \and 
          N. Jackson \inst{\ref{inst:UK}}
          \and
          B. Adebahr \inst{\ref{inst:mpifr}}
          \and 
          M.~A. P\'erez-Torres \inst{\ref{inst:iaa},\ref{inst:cefca},\ref{inst:unizar}}
          \and 
          K. T. Chy\.zy \inst{\ref{inst:chyzy}}
          \and
          T. D. Carozzi \inst{\ref{inst:chalmers}}
		  \and
		  J. Moldón \inst{\ref{inst:ASTRON}}
          \and 
          S. Aalto \inst{\ref{inst:chalmers}}
          \and 
          R. Beck \inst{\ref{inst:mpifr}}
          \and
          P. Best \inst{\ref{inst:SUPA}}
          \and
          R.-J. Dettmar \inst{\ref{inst:bochum}}
          \and\\
          W. van Driel \inst{\ref{inst:Wim}}
          \and
          G. Brunetti \inst{\ref{inst:INAF}}
          \and
          M. Br\"uggen \inst{\ref{inst:hamburg}}
          \and
          M. Haverkorn \inst{\ref{inst:radboud}, \ref{inst:leiden}}
          \and 
          G. Heald \inst{\ref{inst:ASTRON},\ref{inst:kapteyn}}
          \and
          C. Horellou \inst{\ref{inst:chalmers}}
          \and
          M. J. Jarvis \inst{\ref{inst:keble}, \ref{inst:cape}}
          \and\\
          L. K. Morabito \inst{\ref{inst:leiden}}
          \and
          G. K. Miley \inst{\ref{inst:leiden}}
          \and
          H. J. A. R\"ottgering \inst{\ref{inst:leiden}}
          \and
          M. C. Toribio \inst{\ref{inst:ASTRON}}
          \and 
          G. J. White \inst{\ref{inst:open},\ref{inst:RALSpace}}
}

   \institute{
              Department of Earth and Space Sciences,
              Chalmers University of Technology, 
              Onsala Space Observatory,
              439 92 Onsala, 
              Sweden \\
              \email{varenius@chalmers.se}
              \label{inst:chalmers}
              \and
              Jodrell Bank Centre for Astrophysics,
              Alan Turing Building,
              School of Physics and Astronomy,
              The University of Manchester,
              Manchester M13 9PL,
              UK
              \label{inst:UK}
              \and
              The Netherlands Institute for Radio Astronomy (ASTRON), PO Box 2, 7990 AA Dwingeloo, The Netherlands
              \label{inst:ASTRON}
              \and
              Max-Planck-Institut f\"ur Radioastronomie, Auf dem H\"ugel 69, D-53121 Bonn, Germany
              \label{inst:mpifr}
              \and
              Instituto de Astrof\'isica de Andaluc\'ia, Glorieta de las Astronom\'ia, s/n, E-18008 Granada, Spain.
              \label{inst:iaa}
              \and
              Centro de Estudios de la F\'isica del Cosmos de Arag\'on, E-44001 Teruel, Spain.
              \label{inst:cefca}
              \and
              Departamento de F\'isica Teorica, Facultad de Ciencias, Universidad de Zaragoza, Spain.
              \label{inst:unizar}
              \and
              Obserwatorium Astronomiczne Uniwersytetu, Jagiello\'nskiego, ul. Orla 171, 30-244 Krak\'ow, Poland
              \label{inst:chyzy}
              \and 
              SUPA, Institute for Astronomy, Royal Observatory Edinburgh, Blackford Hill, Edinburgh, EH9 3HJ, UK
              \label{inst:SUPA}
              \and
              Ruhr-Universit\"at Bochum, Astronomisches Institut, 44780 Bochum, Germany
              \label{inst:bochum}
              \and
              GEPI, Observatoire de Paris, CNRS, Universit\'e Paris Diderot, 5 place Jules Janssen, 92190 Meudon, France
              \label{inst:Wim}
              \and
              INAF-Istituto di Radioastronomia, via P. Gobetti 101,I--40129 Bologna, Italy
              \label{inst:INAF}
              \and
              Hamburger Sternwarte, University of Hamburg, Gojenbergsweg 112, 21029 Hamburg, The Germany
              \label{inst:hamburg}
              \and
              Department of Astrophysics/IMAPP, Radboud University, P.O. Box 9010, 6500 GL Nijmegen, The Netherlands
              \label{inst:radboud}
              \and
              Leiden Observatory, Leiden University, P.O. Box 9513, NL-2300 RA Leiden, the Netherlands
              \label{inst:leiden}
              \and
              Kapteyn Astronomical Institute, Postbus 800, 9700 AV, Groningen, The Netherlands
              \label{inst:kapteyn}
              \and
              Astrophysics, Department of Physics, Keble Road, Oxford OX1 3RH, UK
              \label{inst:keble}
              \and
              Department of Physics, University of the Western Cape, Private Bag X17, Bellville 7535, South Africa
              \label{inst:cape}
              \and
              Department of Physical Sciences, The Open University, Milton Keynes MK7 6AA, England, and 
              \label{inst:open}
              \and
              RALSpace, The Rutherford Appletion Laboratory, Chilton, Didcot, Oxfordshire OX11 0NL, England
              \label{inst:RALSpace}
    }
   \date{Received 1 October 2014 / Accepted 26 November  2014}

 
  \abstract
   {The nuclear starburst in the nearby galaxy M82 provides an excellent
       laboratory for understanding the physics of star formation. This galaxy
       has been extensively observed in the past, revealing tens of radio-bright
       compact objects embedded in a diffuse free-free absorbing medium.  Our
       understanding of the structure and physics of this medium in M82 can
       be greatly improved by high-resolution images at low frequencies where
       the effects of free-free absorption are most prominent. 
}
   {The aims of this study are, firstly, to demonstrate imaging using international
baselines of the Low Frequency Array (LOFAR), and secondly, to constrain low-frequency
spectra of compact and diffuse emission in the central starburst region of M82
via high-resolution radio imaging at low frequencies.}
   {The international LOFAR telescope was used to observe M82 at 110-126\,MHz
and 146-162\,MHz. Images were
obtained using standard techniques from very long baseline interferometry.
images were obtained at each frequency range: one only using international
baselines, and one only using the longest Dutch (remote) baselines.  
}
{The 154\,MHz image obtained using international baselines is a new imaging record
in terms of combined image resolution (0.3$''$) and sensitivity
($\sigma$=0.15\,mJy/beam) at low frequencies ($<327$\,MHz). We detected 16
objects at 154\,MHz, six of these also at 118\,MHz.  Seven objects detected at
154\,MHz have not been previously catalogued.  For the nine objects previously detected,
we obtained spectral indices and emission measures by fitting models to spectra
(combining LOFAR with literature data).  Four weaker but resolved features are also
found: a linear (50\,pc) filament and three other resolved objects, of which two
show a clear shell structure.  We do not detect any emission from either
supernova 2008iz or from the radio transient source 43.78+59.3.  The images
obtained using remote baselines show diffuse emission, associated with the
outflow in M82, with reduced brightness in the region of the edge-on star-forming disk. 
}
   {}
   \maketitle
%

\section{Introduction}
The nearby \citep[3.52$\pm$0.02\,Mpc;][]{tully2009} galaxy M82 is one of the
most studied extragalactic sources across the electromagnetic spectrum.  In the
radio band, M82 has been observed in a wide range of spatial resolutions in both
the continuum and the spectral line. Multiple radio-emitting compact objects were first
discovered and confirmed in the nucleus more than 30 years ago
\citep{kronberg1972,unger1984}.  To date, high-resolution continuum observations
have revealed more than 50 compact objects thought to be radio supernovae
(SNe), supernova remnants (SNRs) or HII-regions \citep{gendre2013}.  The M82
nucleus provides an excellent laboratory for studying the physics of these
compact objects and the internal medium in which they are embedded.  
Optical observations \citep{westmoquette2009} show that the ionised gas in the
starburst core of M82 is dynamically complex.  The properties of the absorbing
ionised gas component in M82 can be studied by its low-frequency free-free
absorption of continuum emission. A key motivation of such observations is to
determine the structure of the absorbing medium; uniform or ``clumpy''
\citep{lacki2013}. 
Evidence of clumpy free-free absorption has been found
in many galaxies using low-frequency observations \citep{israel1990,israel1992}.
Detailed modelling of SNR evolution at low frequencies is also important
for improving understanding of cosmic ray injection in galaxies.

The integrated low-frequency radio spectrum of M82 has been studied for more than
20 years \citep{condon1992}.  Recently, M82 has been imaged at 327\,MHz
($\lambda=92$\,cm) with 40$''$ resolution using the Westerbork synthesis radio 
telescope (WSRT) by \cite{adebahr2013}, observations 
that show a radio-bright extended halo structure for M82.  However, 
imaging with resolution of tenths of an arcsecond is required 
to properly study the compact objects at low frequencies.

Pioneering subarcsecond observations were done in the 1960s using very long
baseline interferometry to probe angular scales as small as 0.1$''$ in bright
objects such as Jupiter (e.g. \citealt{carr1970,Clark1975}).  However,
these observations were limited to measuring sizes (or upper
limits) using visibility amplitude information and no imaging was performed. 

Imaging with subarcsecond resolution has been carried out at 327\,MHz by,
amongst others, \cite{wrobel1986}, \cite{ananthakrishnan1989} and
\cite{lenc2008}, using international and intercontinental baselines.  At lower
frequencies, imaging has been performed with arcsecond resolution, for example
using the 151\,MHz ($\lambda=$2\,m) system installed on the Multi-Element Radio
Linked Interferometer Network (MERLIN) in the mid, to, late 1980s
\citep{leahy1989}, which had baselines up to 217\,km but could only record a
narrow bandwidth (typically 1\,MHz).  

Low-frequency turnovers in SNR spectra have been used to study the clumpy
absorbing medium in our galaxy in detail (see e.g.  \citealt{kassim1989}).
Pioneering very high resolution (0.5$''$) observations of M82 by
\cite{wills1997} using MERLIN at 408\,MHz ($\lambda=74$\,cm) have shown
evidence of low-frequency turnovers in the spectra of multiple compact objects
in M82 which is mainly attributed to free-free absorbing gas.

However, for many objects, the
low-frequency spectra are not well constrained, and some may have spectral
turnovers well below 408\,MHz. To investigate such objects, as well as to
potentially detect new classes of very steep spectrum compact sources, requires
high-resolution observations at lower frequencies.  Achieving a few tenths of
an arcsecond resolution, as required to separate compact sources from diffuse
emission, needs 1000\,km baselines and hence the international baselines of the
Low Frequency Array (LOFAR).  International LOFAR baselines are only sensitive 
to very compact objects, which in M82 are weaker than the strong
($>$10\,Jy) extended radio emission seen at shorter baselines towards the
centre of the galaxy.  The above scientific goals and its high declination
make M82 an excellent target for demonstrating international LOFAR imaging on a
weak but complex target
source. 
Observations using the international baselines of LOFAR are not yet a regularly
used mode. Pioneering work using the low band (30-80\,MHz) 
succeeded in imaging the source 3C196 with a resolution of $\sim 1''$ only using
 a fraction of the final array \citep{wucknitz2010}.  This commissioning 
continued in the high band (110-190\,MHz), where a resolution of $0.3''$ was
achieved for a number of bright and compact sources \citep{wucknitz2010_2}. 
The LOFAR community have concluded that standard very long baseline
interferometry (VLBI) techniques can be used to produce high-resolution images
from LOFAR observations. Several unpublished high-band observations of Jansky
level sources have been made and multiple observations including international
LOFAR baselines have been scheduled in LOFAR cycles 0, 1 and 2.

In this paper, we present subarcsecond images of the nuclear starburst in M82
at central frequencies of 118\,MHz ($\lambda=2.5$\,m) and 154\,MHz
($\lambda=1.9$\,m), made using international LOFAR baselines, thereby setting a new record in
terms of combined image resolution and point source sensitivity for science
images at frequencies below 327\,MHz.  The contents of this paper are as
follows: in Sect. \ref{sect:cal} we describe in detail the observational setup
and calibration procedures applied to the LOFAR data.  This section also briefly describes
the eMERLIN data used for comparison throughout this paper.  
In Sect. \ref{sect:imaging}, we describe how images were obtained from
LOFAR data and summarise the calibration and imaging procedures.  In Sect.
\ref{sect:randd}, we present the images obtained and briefly discuss our results
in relation to previous work. Finally, we summarise our conclusions in Sect.
\ref{sect:conclusions}.  
A more extensive scientific discussion regarding the compact and extended
emission in M82 will be the subject of a forthcoming paper.  
Throughout this paper we adopt a distance to M82 of 3.52\,Mpc
\citep{tully2009} so that $1''$ corresponds to 17\,pc.


\section{Data and calibration} 
\label{sect:cal}
In this section, we describe the two data sets used in this paper and
the processing done for obtaining images. In particular, we describe in detail
the processing done to image the international LOFAR baselines, since our
strategy differs from what is usually done to image shorter LOFAR baselines.
 
\subsection{eMERLIN data}
The eMERLIN data set was observed on January 20th 2014 at 1.6\,GHz, under
project code CY1203 (P.I.: Pérez-Torres), with additional observations taken in
collaboration with the LeMMINGs e-MERLIN legacy project (P.I.: Beswick \&
McHardy).  These data were calibrated and imaged in a standard way by eMERLIN
staff. Since these data are only used for comparison in this paper, we refer
any details on the calibration to the paper describing these data, see
\cite{miguel2014}.

\subsection{LOFAR data}
Our project LC0\_026 (P.I.: J.E. Conway) was observed in two parts to maximise
hour angle coverage during night time: 10 hours taken during the night between
the 20, and 21, March 2013 (in UT range 16:15-02:45) and six hours taken
in the evening of April 5, 2013 (in UT range 17:45-00:15 UT).  Both the March and
April observations included the same 44 LOFAR high band antenna (HBA) stations:
23 core stations (CS), 13 remote stations (RS), and eight international (INT)
stations. Participating INT stations were DE601, DE602, DE603, DE604, DE605,
FR606, SE607, UK608, although no fringes were detected to DE604.

This experiment was designed to observe, as targets, both the AGN in M81
(hereafter M81*) and the nucleus of the galaxy M82; these two objects are
only separated by 0.61$^\circ$ on the sky (scientific results on M81* will be
presented in a future paper).  Three objects, J0958+6533 (hereafter J0958),
M81* and M82, were observed simultaneously using three beams. Even though M81*
and M82 lie within the width of a single HBA station beam, the field of view
(FOV) of international baseline observations is limited by frequency and time
averaging effects (see below), so that two separate correlation phase centres
were needed.  The source J0958, at angular distance of
3.5$^\circ$ from M81* and 4.1$^\circ$ from M82, was used as a calibrator of
fringe-rate and delay.  This source is listed in the VLBA calibrator list as a
compact source and was known to have a flux density of 0.86\,Jy at 73.8\,MHz in
VLSSr \citep{lane2014}. 
M81* was observed both as a scientific target
and as a close phase calibrator for M82.  Every hour, the observations switched
to a single beam on 3C196 for two minutes.  3C196 was observed to anchor the
absolute flux scale of the observations.  The positions assumed for correlation
and calibration of each source are listed in Table \ref{tab:targetlist}.

Data were taken in 8-bit HBA joined mode, where the data from each HBA subfield
(or ``ear'') in a CS are combined at the station level. The available total
bandwidth of 96\,MHz was divided equally between the three beams on M81*, M82
and J0958.  The single beam on 3C196 covered the same 32\,MHz.  The observed
32\,MHz bandwidth per beam was divided into two contiguous blocks of 16\,MHz
each, one centred on 118\,MHz and one centred on 154\,MHz.  The lower frequency
block was chosen to include the lowest frequencies observable with the HBA, and the higher
frequency range was placed in the region with the highest HBA sensitivity. The
two blocks were calibrated separately but following the same procedure.

The data were correlated producing all four linear polarisation products (XX,
XY, YX, YY).  After correlation, the data were stored in the LOFAR long term
archive (LTA) as measurement sets (MS) with integration time 1.0\,sec. and 64ch./sub-band, 
corresponding to a channel resolution of 3\,kHz. 
In addition to the raw high-resolution data, the LOFAR pipeline was used to
automatically edit bad data, and save the edited data set averaged in time and
frequency down to 2\,sec. integration time and 4ch./sub-band (giving a channel
width of 48\,kHz).  These averaged data were saved in the LOFAR long term
archive (LTA). Inspection of data for the bright calibrator J0958 revealed
that ionospheric phase errors varied slowly enough over time and frequency to
enable further averaging without significant coherence losses.  To speed up the
processing the data were therefore finally averaged, using the LOFAR \emph{new default
pre-processing pipeline} (NDPPP), to 10.0\,s integration time and 1\,ch./sb.
(channel width 195\,kHz). At this point the total data volume was reduced to
170\,GB for each of the two frequency blocks.  

\subsubsection{Baseline definitions and coherence losses}
The international LOFAR telescope provides a very good sampling of Fourier
space covering baselines of lengths between 0.1\,k$\lambda$ and 550\,k$\lambda$
(except for a gap between 35\,k$\lambda$ and 65\,k$\lambda$) at 118\,MHz, and
between 0.1\,k$\lambda$ and 700\,k$\lambda$ (except for a gap between
45\,k$\lambda$ and 80\,k$\lambda$) at 154\,MHz\footnote{This gap will be
largely filled when the final Dutch remote stations are operational}.  In this
paper we present images made using Dutch (remote) baselines and international
baselines.  At these frequencies, remote baselines are baselines of length between
2\,k$\lambda$ and 60\,k$\lambda$.  International baselines are defined as
longer than 60\,k$\lambda$.  

Given the final spectral and temporal resolution of the data (10.0\,sec. 1ch./sb.),
we estimate the coherence loss due to time and frequency smearing at an angular
distance of $30''$ from the phase centre to be less than 3\% for the longest
international (1158\,km) baseline.
This estimate includes a simple upper bound for time smearing assuming
the source to be at the celestial north pole.  Similarly, we estimate the
coherence loss to be less than 3\% for the longest Dutch remote baselines
(121\,km) for emission at an angular distance of $5'$ from the centre.
Including residual rates and delays (typically 3\,mHz and 300\,ns) present
when averaging the data we estimate a total coherence loss of less than 5\%.
Hence, we may image a field of $1'$ around the central M82 position using INT
baselines and $10'$ using RS baselines, without significant loss due to
averaging.

   \begin{table}
      \caption[]{List of correlation positions for each beam.}
         \label{tab:targetlist}
         \begin{tabular}{ l l r}
            Source      &  R. A. [J2000] & Dec. [J2000]\\
            \hline
            3C196 \tablefootmark{a}& 08h13m36.0000s & 48$^\circ13'03''$.000\\
       J0958+6533 \tablefootmark{b}& 09h58m47.2451s & 65$^\circ33'54''$.818\\
              M81* \tablefootmark{c}& 09h55m33.1731s & 69$^\circ03'55''$.062\\
              M82 \tablefootmark{d}& 09h55m51.5500s & 69$^\circ40'45''$.792\\
            \noalign{\smallskip}
            \hline
         \end{tabular}
         \tablefoot{
             \tablefoottext{a}{From NED (http://ned.ipac.caltech.edu).}
             \tablefoottext{b}{From the VLBA Calibrator list (http://www.vlba.nrao.edu/astro/calib).}
             \tablefoottext{c}{Centred on the core M81* \citep{bietenholz2004,martividal2011}.}
             \tablefoottext{d}{Centred on supernova SN2008iz in M82 \citep{brunthaler2009}.}
}
   \end{table}

\subsubsection{Correcting for residual delay, rate and phase}
\label{sec:fring}
Because of residual rates affecting international baselines we need to derive
rate corrections using a global fringe-fitting algorithm. This has not yet been
implemented within the LOFAR software packages, nor in the Common Astronomy
Software Applications (CASA) 4.2.1 \citep{CASA}.  Therefore, we decided to use
Astronomical Image Processing System (AIPS) 31DEC13 \citep{greisen} and
ParselTongue 2.0 \citep{Kettenis}, to calibrate these data.

Distortion of the signal due to the ionosphere is challenging to remove in a
linear (X, Y) polarisation basis, since both amplitudes and phases are affected
in a coupled way.  To simplify the problem, we convert to a circular (R,L) polarisation basis,
where the ionospheric disturbance is transformed to phase-only effects, and
employ standard VLBI techniques.  Since Faraday rotation does not mix R and L
polarisations we may calibrate RR and LL independently.
Hence, in a circular basis, corrections for phase and amplitude can be derived
independently. The data were converted from linear to circular
polarisation using the tool \emph{mscorpol} v1.6, developed by T.  Carozzi.
This tool includes corrections for dipole-projection effects as a function of
the correlated sky position relative to all included LOFAR stations.  After the conversion,
the data are circularly polarised, with full (but approximated) parallactic angle
correction.  The data were then converted to UVFITS format using the task
\emph{exportuvfits} in CASA and read into AIPS.

Before further processing, the task FIXWT was used
to set the relative weights of all visibilities to the inverse square of the
standard deviation within five minutes of data. Hence, the relative weights should
reflect the scatter of the data.

The global fringe fitting algorithm \citep{thompson}, as implemented in the
AIPS task FRING, only solves for a single residual delay within each
intermediate frequency (IF).  At low radio frequencies, the residual delay due
to the ionosphere varies considerably as a function of frequency.  Since these
data cover a large fractional bandwidth ($10\%$), solving for one single delay
over the full 16\,MHz might leave significant residual delays in parts of the
spectrum.  To mitigate this effect, each 16\,MHz block (of 81 channels) was
split in three groups (``IFs'' in AIPS) of 27 channels each, using the task
MORIF.

Corrections for residual delays and rates were derived using J0958.  
The search was restricted to baselines longer than 60k$\lambda$, a delay search
window of 600\,ns, a rate search window of 30\,mHz, and a solution interval of
two minutes.  Solutions were found separately for each IF and polarisation.
Since all CS are close and share a common clock, no corrections were needed for
the core stations, only for the INT and RS.  Typical residual delays were found
to be 100-300\,ns, and typical residual rates were 1-3\,mHz. 
To correct a few obvious outliers, the corrections were filtered using a median 
window filter before applying them to the data.
The difference in residual delay between the lowest and highest frequencies of the three
IFs were typically 10-15\,ns for the international stations.  Bad data were
edited using UVFLG in AIPS.  No fringes were detected to DE604 and this station
was not used in the imaging. We note that the problems with DE604 have been
fixed since these data were taken.

\subsubsection{Setting the absolute flux scale}
\label{sect:3C196}
After correcting for residual delays and rates, the task CALIB was used to derive
amplitude and phase corrections for J0958, assuming this source to be a 0.5\,Jy
flat-spectrum point source at the phase centre.  A solution interval of two 
minutes was used together with the ``L1-norm'' option, and the two circular
polarisations were averaged to form a single solution for both polarisations.

\begin{figure}[htbp]
\centering
\includegraphics[width=0.48\textwidth]{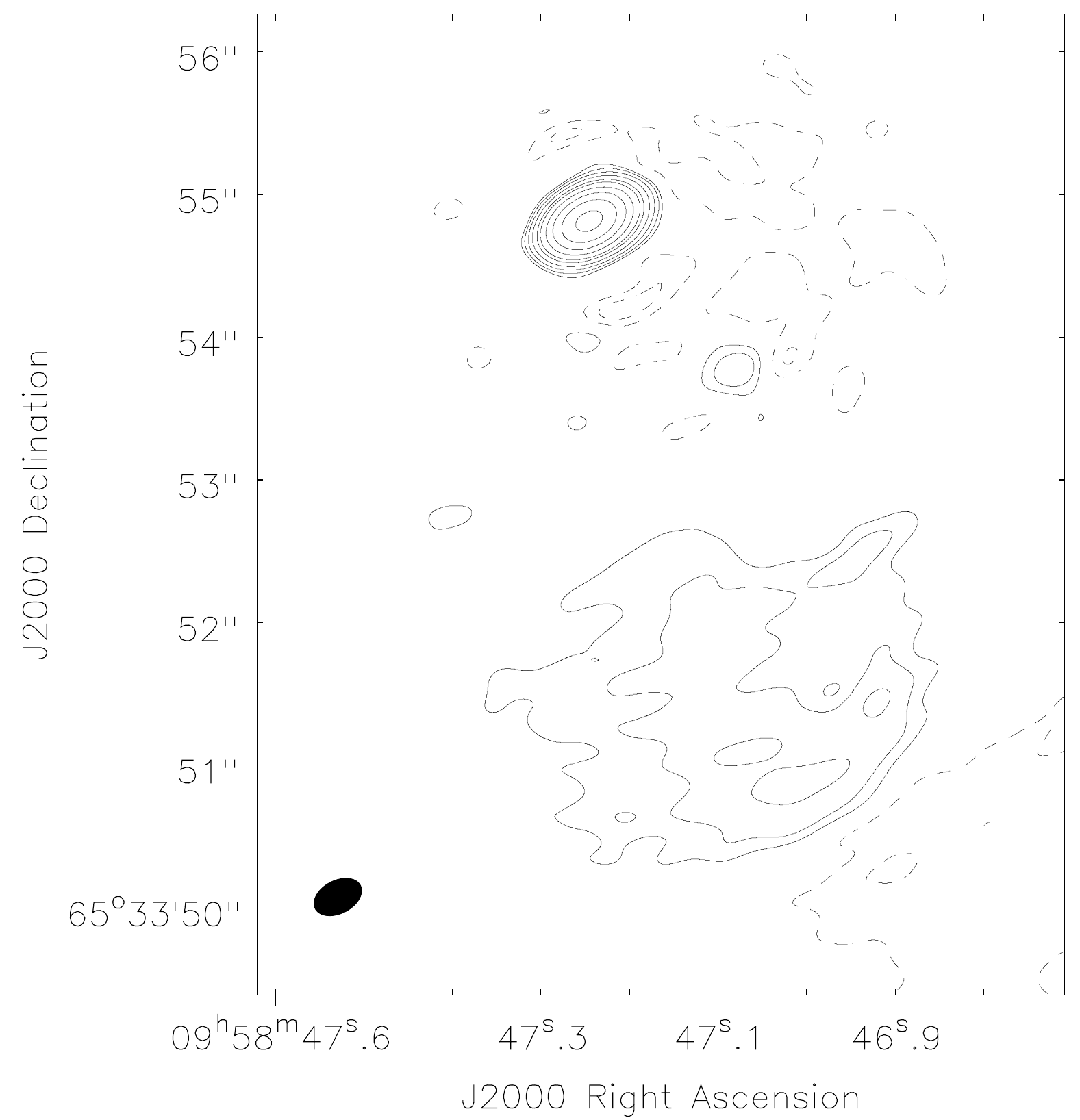}
\caption{
Image of the calibrator J0958+6533 at 154\,MHz using baselines lonnger than
$60\mathrm{\,k}\lambda$.  The RMS noise is $\sigma$=0.2\,mJy/beam.  The beam of
$0.36''\times0.23''$ is shown in the bottom left. The contour levels are (-10, -5, 5, 10,
20, 40, 80, 125, 250, 500, 1000, 2000)$\times \sigma$.  A southwest extension
is clearly visible, although the point source is 100 times brighter than the
extension.  This image was obtained using the multi-scale option in CLEAN (CASA
4.2.1) with scales = [0, 40, 80, 160] pixels, with pixel size $0.02''$.
}
    \label{fig:J0958}
\end{figure}

J0958 is compact, but there is another strong (0.5\,Jy) source nearby (distance
$4.7'$ north of J0958).
Only baselines $>60k \lambda$ were used when determining amplitude corrections,
thereby limiting the field of view around J0958 because of smearing, which
means avoiding any disturbing interference from this source.
For clarity we emphasise
that the amplitude corrections are derived for all stations, INT, RS, and CS,
but only using the longest baselines.  The solutions and final corrected
visibilities were inspected to ensure that good solutions were found.  After
applying the corrections, imaging using task IMAGR in AIPS recovered
a compact source of 498\,mJy at 118\,MHz and 502\,mJy at 154\,MHz (using
baselines $>60k \lambda$, pixel size $0.02''$ and robust 0 weighting;
) in good agreement with what was specified. Flux densities
are given as reported by fitting Gaussian intensity distributions 
using JMFIT in AIPS.  The CLEAN restoring beam was
$0.48''\times0.33''$, RMS noise $\sigma$=0.4\,mJy/beam at 118\,MHz, and
$0.36''\times0.23''$, RMS noise $\sigma$=0.2\,mJy/beam at 154\,MHz.  In addition to
the compact source, a weak 5$''$ extension can be seen to the southwest (see
Fig. \ref{fig:J0958} made using the multi-scale option in CASA to better
deconvolve extended emission).  This southwest extension is known from previous
observations at 1.4\,GHz by \cite{xu1995}.  Since the point source is 100
times brighter than the extension, a point source model is sufficiently
accurate to determine calibration solutions.  In case the extended structure
has introduced minor phase errors, these will be corrected using M81* before
imaging of M82. 

Imaging using baselines between 2 and $60k \lambda$ (i.e. excluding baselines
to INT stations as well as CS-CS) we recover 686\,mJy with a peak of
592\,mJy/beam at 118\,MHz (CLEAN beam $6.35''\times5.37''$, pixel size $0.5''$,
$\sigma=0.33\mathrm{\,mJy/beam}$) and 646\,mJy with a peak of 549\,mJy/beam at 154\,MHz
(CLEAN beam $5.07''\times4.34''$, pixel size $0.5''$, $\sigma=0.14\mathrm{\,mJy/beam}$).
It is clear that J0958 is partially
resolved at these frequencies and the integrated flux density recovered is
hence a lower limit. We note that the integrated flux density recovered at
73.8\,MHz of 860\,mJy  \citep[VLSSr beam $75''$]{VLSS} is 25\% higher than found
here at 118\,MHz.

To check the flux scale, the cumulative corrections derived for J0958 were
applied to 3C196. Because of the large angular separation (22$^\circ$) of 3C196
with respect to J0958, the phase calibration was refined for 3C196 using the CS
and RS, assuming a point source model and deriving one solution for each 2\,min
scan.  The corrections were found using baselines in the range 0.1 to
$60k\lambda$ (excluding the shortest CS baselines as well as INT baselines).
The two polarisations were averaged together, but each IF was solved
separately.  

We note that 3C196 is resolved at the RS baselines, but still compact enough
for a point source model to work for phase calibration.  Because of the limited
Fourier sampling (16$\times$2\,min) we did not calibrate and image the INT
station visibilities of 3C196.  However, this was not necessary to obtain the
integrated flux densities at 118 and 154\,MHz.  For clarity we emphasise that
no corrections derived for 3C196 were transferred to any other source; 3C196
was just imaged to check the amplitude calibration.  The amplitude corrections
for all stations were derived using J0958. 
We also note that no beam corrections were made in addition to what was
applied by \emph{mscorpol}. Because of the small field of view imaged 
in this work, any residual beam errors should have a very minor impact.
This calibration strategy is different (and simpler) than what is normally used
when imaging wide fields (where the field of view is comparable to the station beam) with LOFAR.

Imaging of 3C196 was done using baselines of length 0.1-60\,k$\lambda$, pixel size
$0.2''$ and robust 0 weighting.  The resulting images recover 102.0\,Jy at
118\,MHz (CLEAN beam $7.37''\times5.40''$, $\sigma=\mathrm{0.3\,Jy/beam}$) and 84.2\,Jy
at 154\,MHz (CLEAN beam $7.57''\times5.95''$, $\sigma=\mathrm{0.1\,Jy/beam}$). 

We estimate our absolute flux calibration to be accurate to within 10\%.  With
this in mind, the recovered flux densities for 3C196 are in excellent agreement
with the 98.0\,Jy and 81.6\,Jy expected at 118\,MHz and 154\,MHz respectively
from the best-fit model for 3C196 derived by \cite{scaife2012}. 

\subsubsection{Refining the phase calibration using M81*}
\label{sect:refine}
We found M81* to be too weak to determine delay and rate corrections using FRING.
Approximating the ionosphere as a simple slab, and assuming a typical observing
elevation of 60$^\circ$ and a typical delay in direction of J0958 as 200\,ns,
we estimate the delay difference over the 3.5$^\circ$ angular separation on the sky between J0958
and M82* to be 10\,ns. This would, given the full 16\,MHz bandwidth, introduce
a coherence loss of 1.8\% when averaging all channels for M81 and M82.
We find this acceptable and cumulative corrections derived for J0958 (delay,
rate, amplitude and phase) were transferred to M81*.  Because of the angular
separation between J0958 and M81* it was necessary to do phase-only calibration
on M81* to derive additional phase corrections towards M81*.  A point-source
model was used and corrections were derived using baselines $>60k\lambda$,
averaging over IFs and polarisation.  At 118\,MHz corrections were derived
every four minutes, and at 154\,MHz corrections were derived every two minutes.

Deconvolved images of M81* were obtained using the AIPS task IMAGR.  Using baselines
between 2 and $60k \lambda$ and pixel size $0.5''$ we recover 67.9\,mJy with a
peak of 52.1\,mJy/beam at 118\,MHz (CLEAN beam $6.22''\times5.17''$
$\sigma=\mathrm{0.37\,mJy/beam}$), and 68.6\,mJy with a peak of 52.7\,mJy/beam at 154\,MHz
(CLEAN beam $4.92''\times4.05''$, $\sigma=\mathrm{0.16\,mJy/beam}$). Since the peak flux
density is smaller than the itegrated flux density it is clear that M81* is 
partially resolved at both frequencies
using RS baselines.  Most of the flux comes from a compact ($5''$) central
source. There is a hint of emission extending up to $10''$ from the centre but
this could also be a result of cleaning artefacts.  Using baselines $>60k \lambda$ and
pixel size $0.03''$ we only see a very compact source of flux density
38.0\,mJy at 118\,MHz (CLEAN beam $0.47''\times0.33''$
$\sigma=\mathrm{0.20\,mJy/beam}$), and 42.3\,mJy at 154\,MHz (CLEAN beam
$0.41''\times0.25''$ $\sigma=\mathrm{0.11\,mJy/beam}$).  

The cumulative corrections derived from J0958 and M81* were then applied to the
M82 data.  Since M82 is at angular distance 0.6$^\circ$ from M81* residual
phase errors might be present in the M82 data limiting the dynamic range,
although no clear phase errors are visible in the images. We therefore tried
phase-only self-calibration of M82 deriving one solution every five minutes,
but this did not result in any significant improvement of the final images. We
therefore discarded these corrections, and the images presented in this paper
were obtained without any self-calibration on M82. 
Finally the data were exported from AIPS and converted once more to measurement
set (MS) format using the task \emph{exportuvfits} in CASA. To run CLEAN in
CASA the MS ANTENNA table had to be manually changed to contain valid ``mount
type'' for all antennas. Since we are not doing mosaicking, any valid mount
type will be treated equally in CASA, and we used mount=``X-Y''.

\section{Imaging of M82}
\label{sect:imaging}
Because of the large fractional bandwidth (10\%) and large-scale
emission present \citep[$5'$ at $\lambda=92$\,cm]{adebahr2013}, deconvolution
should be done using multi-scale multi-frequency-synthesis (MSMFS).  We
deconvolved the calibrated M82 data using the MSMFS algorithm v2.6 as
implemented in the task CLEAN in CASA 4.2.1 \citep{conway1990,rau2011}.  A
possible dependence of intensity on frequency was taken into account, using the
MFS algorithm with two Taylor terms, for all images presented here.  Note that no
corrections were made for the station beam or other wide-field effects, apart from the
corrections applied by \emph{mscorpol}. 
However, due to the small field of view when imaging, these
effects are negligible for all the results presented here, and we did not need to 
use wide-field imaging software (such as the AW-imager). 

\subsection{Imaging of compact structure in M82}
\label{sect:compact}
The compact emission was imaged only using baselines longer than $60\mathrm{\,k}\lambda$ 
and a pixel size of $0.02''$. This gave a resolution (CLEAN
beam) of $0.45''\times0.29''$ at 118\,MHz and $0.36''\times0.23''$ at 154\,MHz.
Since only compact structure was present the multi-scale option was not used
here. We found that accounting for a possible intensity gradient with respect
to frequency (using MFS with two Taylor terms to model the spectrum of each
pixel) produced almost identical results compared to neglecting any spectral
gradient (only using one Taylor term).  The dirty Stokes V images are empty
of emission, and from these we estimate the minimum image noise
levels as 0.29\,mJy/beam at 118\,MHz and 0.15\,mJy/beam at 154\,MHz.  

The deconvolved Stokes I images, Fig. \ref{fig:M82INT_BOTH}, have 
RMS noise levels of 0.30\,mJy/beam at 118\,MHz and
0.15\,mJy/beam at 154\,MHz. The fact that the residuals resemble
Gaussian noise, with the same standard deviation in both I and V, 
strongly argue that the images 
are not limited by systematic calibration errors or deconvolution effects.
This is not surprising since INT LOFAR, combined with
the MFS algorithm, provides excellent sampling of Fourier space in this
observation, and because distant interfering sources are decorrelated after
averaging visibilities on long baselines in time and frequency.  Fig.
\ref{fig:M82INT} is, to our knowledge, the highest resolution science image ever
published at this and lower frequencies.  Since the Fourier plane was very well
sampled on the smallest scales, the PSF should not introduce imaging artefacts.
We think that the minor ``artefacts'' present in Fig.  \ref{fig:M82INT} are in
fact due to poorly sampled large scale emission surrounding the compact sources
detected. This problem is also experienced when imaging this region of M82
with eMERLIN at higher frequencies.

\subsubsection{Estimating the thermal image noise}
The theoretical image noise for dual polarisation data using natural weighting
can be estimated using a modified version (only including international
baselines) of the image noise equation given in SKA memo 113 by \cite{skamemo113} as
\begin{multline}
\Delta S[\text{Jy}/\text{beam}]=W(4\delta\nu\delta t)^{-1/2}\Bigg(
\frac{N_{INT}(N_{INT}-1)/2}{S^2_{INT}}+\\
        \frac{N_{INT}N_{RS}}{S_{INT}S_{RS}}+
        \frac{N_{INT}N_{CS}}{S_{INT}S_{CS}}
    \Bigg)^{-1/2},
    \label{eqn:noise}
\end{multline}
where $W$ is an extra weighting factor (see below), $\delta\nu$ is the
bandwidth [Hz], $\delta t$ is the integration time [s], $N_{X}$ is the number
of stations of type $X$, and $S_{X}$ is the system equivalent flux density
(SEFD) of station type $X$ [Jy].  A remote station has been estimated to
have a zenith SEFD $S_{RS}\approx2500$\,Jy at 118\,MHz and $S_{RS}\approx1900$\,Jy 
at 154\,MHz \citep[Fig. 22]{haarlem2013}. In
HBA-joined mode we expect a $S_{CS}=S_{RS}$.  INT stations are twice as big,
hence $S_{INT}\approx0.5S_{RS}$.  In addition, \cite{haarlem2013} estimate that the image noise 
increases by a factor of 1.3 due to time-variable station
projection losses, and with a factor of 1.5 due to the robust
weighting used in this paper (as opposed to natural weighting; \citealt{briggs}).  Including
these factors as $W$ in eq. \ref{eqn:noise} and assuming 16\,h integration time,
16\,MHz bandwidth, $N_{INT}=7$, $N_{RS}=13$, and $N_{CS}=23$, we estimate our
theoretical image noise to be 0.11\,mJy/beam at 118\,MHz and 0.08\,mJy/beam at 154\,MHz. 
It is clear that our measured noise levels are 2.9 times higher than  expected
at 118\,MHz and 1.9 times higher at 154\,MHz.
This discrepancy is not fully understood, but there are a few noise-increasing
factors neglected in the above estimate such as increased station SEFDs when
observing at low elevation. The impact of editing should be minor and is in
principle possible to calculate, but in practice it is hard for this
observation when combining manual and automatic editing procedures and we did
not include losses due to editing in the above estimate.
Furthermore, the internal calibration of the international stations was not
optimal at the time of these observations, thereby increasing $S_{INT}$. The
station calibration has, however, been improved since these data were taken.
We note that, despite these hindrances, our attained ratio of image noise to 
theoretical noise is better than that often attained at lower resolution with 
LOFAR, where dynamic range constraints can be severe.

\subsection{Imaging of extended structure in M82}
\label{sect:extended}
Before imaging the extended structure using RS-RS and RS-CS baselines, the
model obtained through deconvolution of the compact structure (see Sect.
\ref{sect:compact}) was subtracted from the data using the task UVSUB in AIPS.
The extended emission was then imaged only using baselines of length between 2
and $60\mathrm{\,k}\lambda$ and pixel size $0.4''$, giving a resolution (CLEAN
beam) of $5.79''\times4.53''$ at 118\,MHz and $4.65''\times3.55''$ at 154\,MHz.
The multi-scale option was used, and the scales were selected as the geometric
series 0, 20, 40, 80, 160 pixels. The largest scale corresponds to the largest
scale expected for the core of M82 \citep[$\sim1'$]{adebahr2013}.  From imaging of
Stokes V, we estimate the minimum image noise to be 0.28\,mJy/beam at
118\,MHz and 0.15\,mJy/beam at 154\,MHz.  The deconvolved Stokes I images, Fig.
\ref{fig:M82RS_BOTH}, have RMS noise levels of 0.50\,mJy/beam at 118\,MHz
and 0.27\,mJy/beam at 154\,MHz, indicating that we either did not successfully
deconvolve all the emission present, or there are residual calibration errors
limiting the dynamic range.  

At RS baselines we are limited by dynamic range rather than
random noise. This is not surprising, since at RS baselines the signal from
interfering distant cosmic sources is much stronger because of less decorrelation due
to averaging.  To properly deconvolve such interfering sources,
imaging of a larger field of view (possibly with multi-directional calibration)
might be needed. Further improvement of the image fidelity by, for example,
using different deconvolution algorithms (such as the LOFAR software AW-imager)
is however beyond the scope of this work.

We detect a weak (peak 1.96\,mJy/beam) signal in Stokes V at 154\,MHz tracing
the brightest part of the extended emission visible in Fig. \ref{fig:M82RS}.
This signal could be real, but could also be a residual polarisation error.  In
particular, the conversion to circular polarisation introduces a good, but
approximate, parallactic angle correction.  This, plus possible gain
differences and/or phase offsets between X and Y, may cause leakage between the
derived R and L.  A complete polarisation calibration is beyond the scope of
this work, and will only marginally affect the Stokes I measurements
presented here.  

\subsection{Summary of calibration and imaging}
\label{sect:calsum}
For convenience, let us summarise the calibration and imaging procedures
described in the previous sections.  The observing mode used in this paper made
use of a primary calibrator, J0958, 4 degrees distant from M82 to determine
delays and rates and initial phase solutions.  These delay, rate and phase
solutions were then transferred (3.5$^\circ$) to M81* which  then allowed, by
averaging over  longer timescales and frequency ranges, phase-only
self-calibration solutions to be determined toward the much weaker (40\,mJy)
M81* nucleus.  The amplitude scale was set by the observed flux density of
3C196.  Transferring these phase solutions 0.6$^\circ$ to M82 allowed random
noise limited images to be made.

\section{Results and discussion}
\label{sect:randd}
\begin{figure*}[htbp]
\centering
\subfigure[Image of M82 at 118\,MHz using international (INT) LOFAR baselines.]{
    \includegraphics[width=0.98\textwidth]{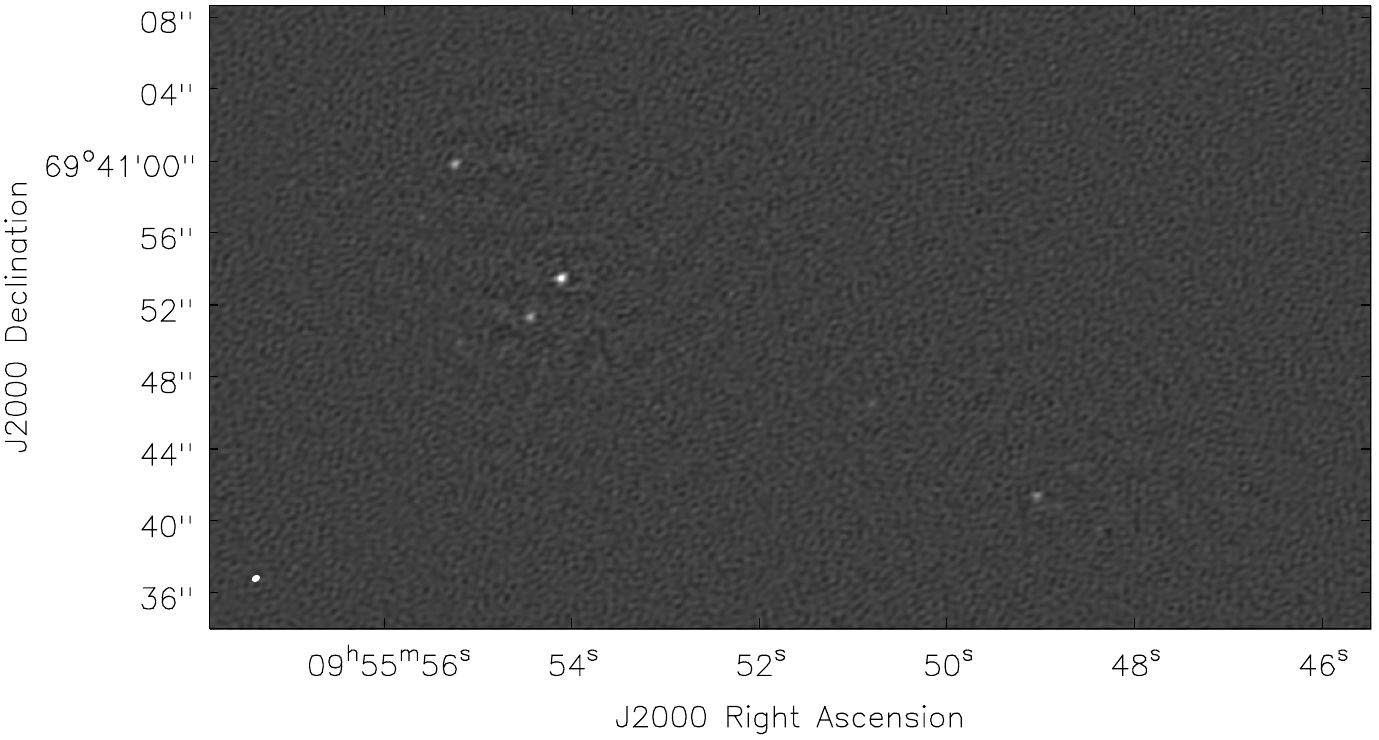}
    \label{fig:M82INT_LOW}
}
\subfigure[Image of M82 at 154\,MHz using international (INT) LOFAR baselines.]{
    \includegraphics[width=0.98\textwidth]{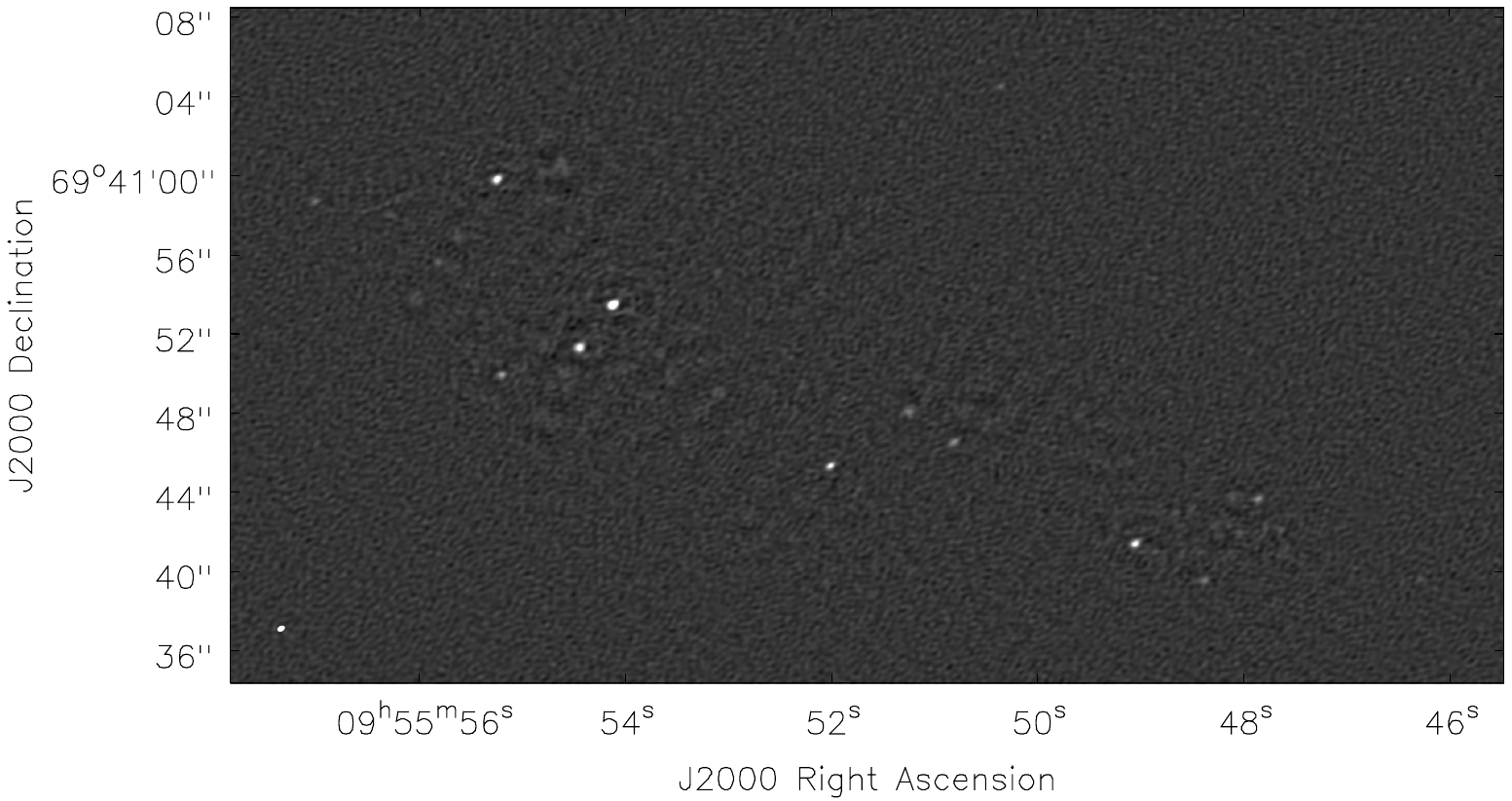}
    \label{fig:M82INT}
}
\caption{
Images of M82 at 118\,MHz (upper panel) and 154\,MHz (lower panel) made using 
international LOFAR baselines (longer than $60\mathrm{\,k}\lambda$). 
The CLEAN restoring beam is shown in the lower left corner of each figure, $0.45''\times0.29''$ at 118\,MHz
and  $0.36''\times0.23''$ at 154\,MHz.
The RMS noise levels are 0.30\,mJy/beam at 118\,MHz and 0.15\,mJy/beam at 154\,MHz. 
For deconvolution parameters, see Sect. \ref{sect:compact}.  Measured
brightness values are listed in Table \ref{tab:props}.
}
\label{fig:M82INT_BOTH}
\end{figure*}

\begin{figure*}[htbp]
\centering
\subfigure[Image of M82 at 118\,MHz using remote (RS) LOFAR baselines.]{
    \includegraphics[width=0.7\textwidth]{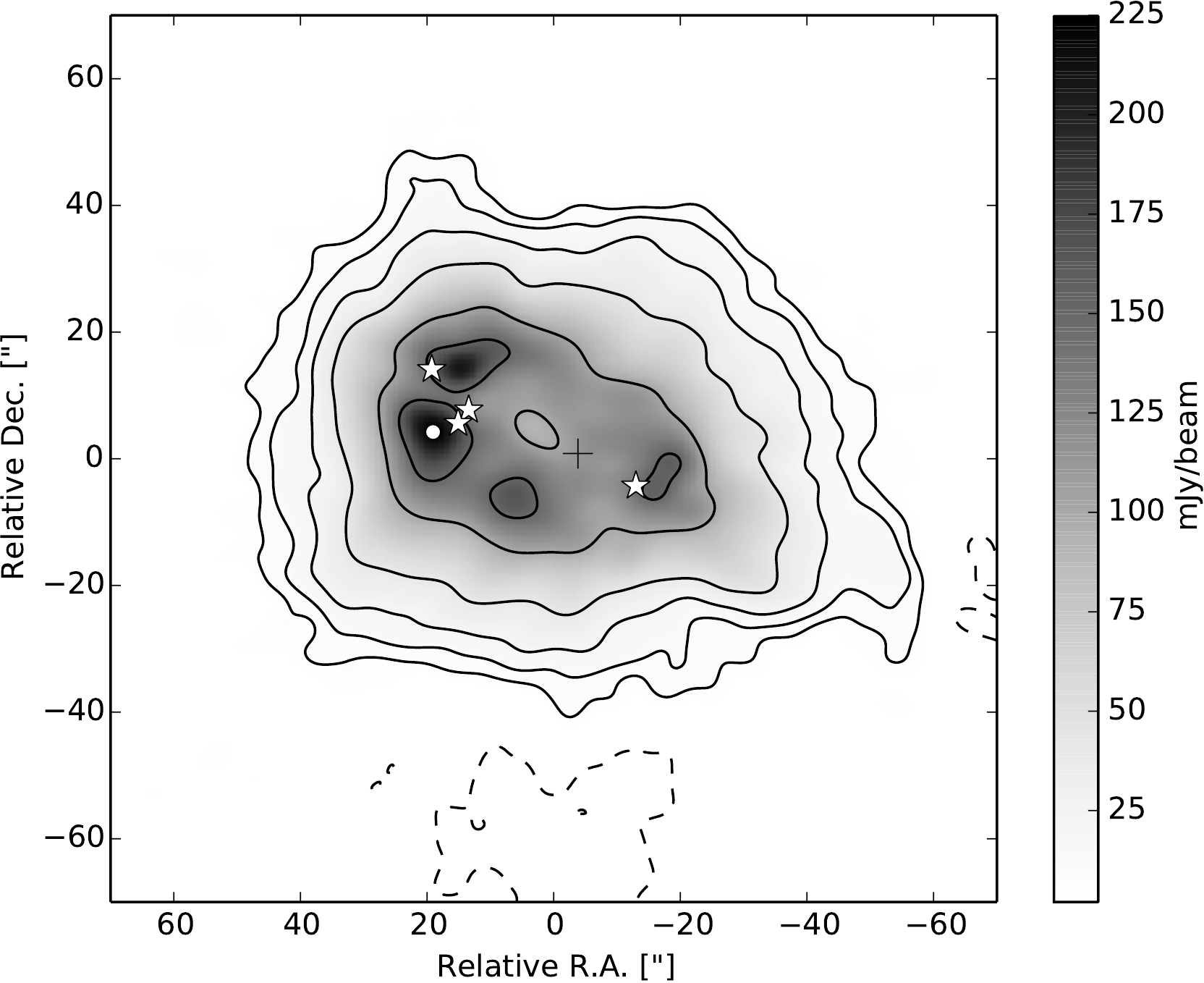}
    \label{fig:M82RSLOW}
}
\subfigure[Image of M82 at 154\,MHz using remote (RS) LOFAR baselines.]{
    \includegraphics[width=0.7\textwidth]{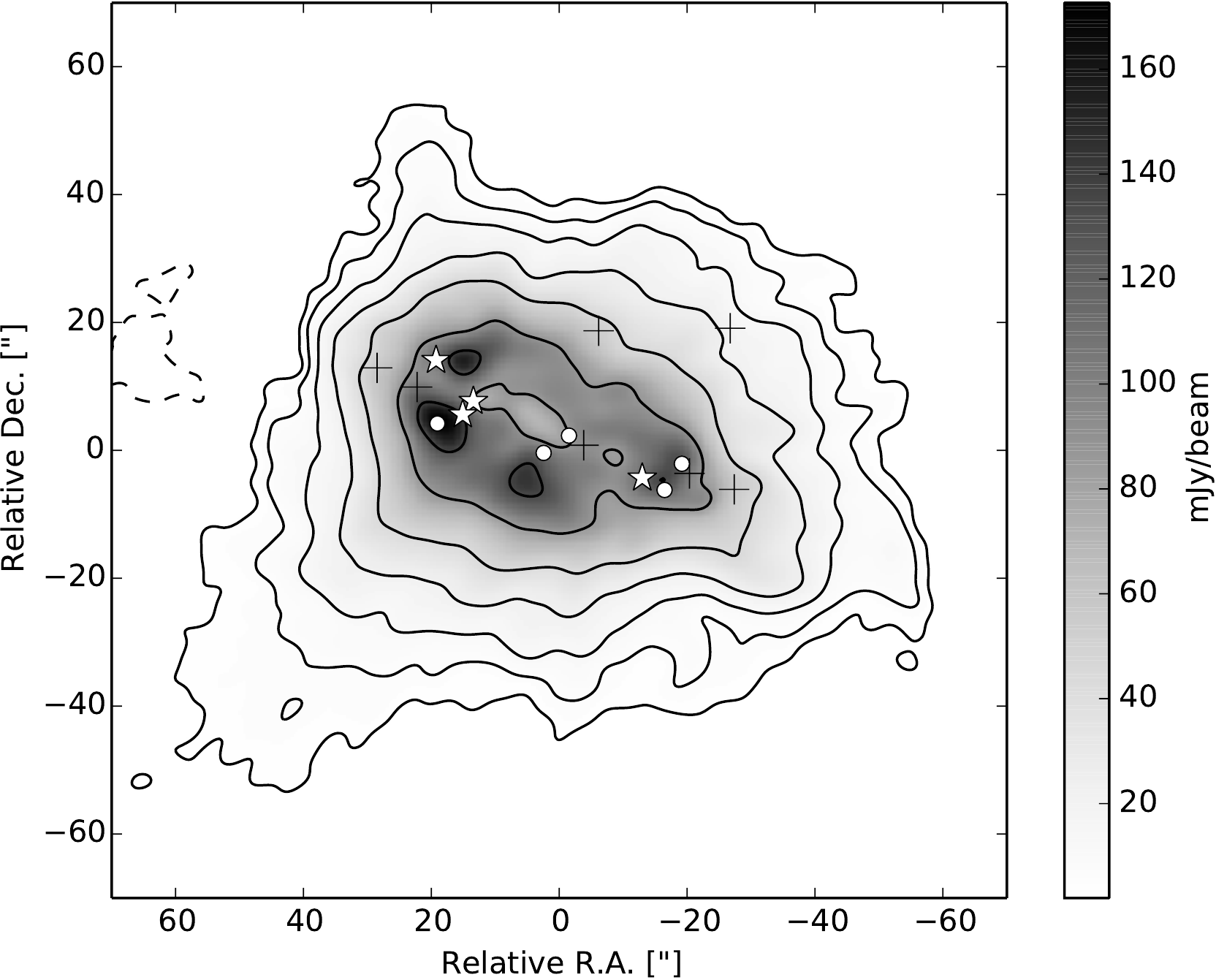}
    \label{fig:M82RS}
}
\caption{
Images of M82 at 118\,MHz (upper panel) and 154\,MHz (lower panel) made using LOFAR baselines of length between
2$\mathrm{\,k}\lambda$ and $60\mathrm{\,k}\lambda$ (robust 0 weighting).  
At 118\,MHz the synthesised PSF (beam) is $5.79''\times4.53''$ and the RMS
noise is $\sigma_{118}=\mathrm{0.50\,mJy/beam}$. At 154\,MHz the PSF is
$4.66''\times3.56''$ and the RMS noise is $\sigma_{154}=\mathrm{0.27\,mJy/beam}$.
For deconvolution parameters, see Sect. \ref{sect:extended}.
These images show the structure of the extended emission in 
grey scale (with brightness per respective beam size). 
The upper panel contours are drawn at (-10, 10, 20, 40, 80, 200, 300)$\times\sigma_{118}$
and the lower panel contours are (-10, 10, 20, 40, 80, 160, 320, 500)$\times\sigma_{154}$.
The symbols mark the positions of the sources listed in Table \ref{tab:props}.
The stars mark the power law-spectrum objects, shown in Fig.
\ref{fig:powerlawspectra}.  The circles mark the turnover-spectrum objects,
shown in Fig.  \ref{fig:turnoverspectra}.  The plus signs mark the
remaining objects listed in Table \ref{tab:props}.  Positions are given
relative to the M82 position in Table \ref{tab:targetlist}.
}
\label{fig:M82RS_BOTH}
\end{figure*}

In this section we present the results along with a brief discussion of the
properties of the compact and diffuse emission seen in Figs.
\ref{fig:M82INT_BOTH} and \ref{fig:M82RS_BOTH}. Fitted quantities for all compact
features detected in Fig. \ref{fig:M82INT_BOTH} are summarised in Table
\ref{tab:props}.  The fitting was done using the software PyBDSM
v1.8.2 with default parameters except but specifying ``hard'' clipping at peak
level of $5\sigma$ and island level of $3\sigma$.  

\subsection{Positional accuracy}
\label{sect:posacc}
Based on the dynamic range in the images we estimate an average positional
uncertainty of $0.04''$ for the objects listed in Table \ref{tab:props}.  When
matching source positions with literature we allowed for an extra $0.05''$
difference since most MERLIN positions were rounded to $0.1''$ in Dec.  We note
that the LOFAR positions appear approximately 60\,mas south-west of the eMERLIN
1.6\,GHz positions.  This offset is visible in Figs. \ref{fig:leftshell} and
\ref{fig:linear}, but is also present for the brighter objects. We believe that
this offset is not due to spectral shifts in any particular object in M82.
In this paper the M82 positions are phase-referenced to M81* assuming the
position for M81* given in Table \ref{tab:targetlist}.  
Because of the 0.6$^\circ$ angular separation between M81* and M82, position
errors may be introduced by slowly varying (i.e. few hours, otherwised averaged when
imaging) differences in ionospheric refraction in the two directions.
Again using the simple slab approximation for the ionosphere, as in Sect. \ref{sect:refine},
we can roughly estimate the residual delay as 1\,ns over 0.6$^\circ$, corresponding to 
a position offset of 60\,mas for a 1000\,km baseline.

In addition to possible ionospheric shifts, we note that M81* is known to have
a core shift at GHz frequencies; the peak of emission is shifted to the
north-east at lower frequencies with respect to images at higher frequencies
\citep{bietenholz2004,martividal2011}, i.e. consistent with the offset seen
between LOFAR and eMERLIN.  Assuming that the power laws, describing the
decrease in the particle density and/or magnetic field as a function of
distance to the jet base, do not change over the whole jet, we estimate, using
the jet model presented by \cite{martividal2011}, the offset due to the
core-shift in M81* to be 10\,mas between 150\,MHz and 1.6\,GHz.  In conclusion,
we find that the offset may be explained by either the ionosphere, or a
core-shift in M81*, or a combination of both. 

   \begin{table*}
\centering
      \caption[]{List of the 16 compact sources detected above $5\sigma$ at
154\,MHz. (Note; four additional possible extended sources are described in
Sect 4.4). Six of the compact sources listed below are detected also at
118\,MHz.  
Fitted peak brightness values are given as $S_P$ and integrated flux densities are given
as $S_I$.  Uncertainties on flux densities were calculated as $((0.1\cdot\{S_I
\text{ or }S_P\})^2+\sigma_{\mathrm{fit}}^2)^{1/2}$ i.e.  including both a 10\%
calibration uncertainty and the uncertainty estimated from the fit,
$\sigma_{\mathrm{fit}}$.  The Right ascension and Declination are given as
offset from $09^h55^m00^s.000$ and 69$^\circ40'00''.00$ (J2000).  
}
         \label{tab:props}
         \begin{tabular}{l r r r r r r r }
             Name & $\Delta$R. A.& $\Delta$Dec. & $S_{P-118}$ & $S_{I-118}$ & $S_{P-154}$ & $S_{I-154}$ \\
             & [s] &[$''$] & [mJy/beam] & [mJy] & [mJy/beam] & [mJy]\\
            \hline
Uncatalogued  &  46.300  &  39.66  &  --  & --  &  0.76$\pm$0.16  & 0.94$\pm$0.26\\
Uncatalogued  &  46.415  &  64.90  &  --  & --  &  0.74$\pm$0.14  & 0.49$\pm$0.27\\
Uncatalogued  &  47.644  &  42.17  &  --  & --  &  0.85$\pm$0.19  & 1.02$\pm$0.30\\
39.10+57.3\tablefootmark{a}\tablefootmark{b}\tablefootmark{c}\tablefootmark{d}  &  47.869  &  43.72  &  --  & --  &  1.94$\pm$0.26  & 2.98$\pm$0.40\\
39.64+53.4\tablefootmark{a}\tablefootmark{b}\tablefootmark{c}\tablefootmark{d}  &  48.389  &  39.61  &  --  & --  &  1.42$\pm$0.23  & 2.30$\pm$0.36\\
40.32+55.1\tablefootmark{a}\tablefootmark{b}\tablefootmark{c}\tablefootmark{d}  &  49.060  &  41.47  &  3.80$\pm$0.51  & 5.92$\pm$0.79  &  5.17$\pm$0.54  & 6.75$\pm$0.73\\
Uncatalogued  &  50.366  &  64.55  &  --  & --  &  1.05$\pm$0.18  & 1.26$\pm$0.27\\
Uncatalogued  &  50.820  &  46.60  &  1.68$\pm$0.36  & 2.29$\pm$0.56  &  1.92$\pm$0.26  & 3.15$\pm$0.41\\
42.53+61.9\tablefootmark{a}  &  51.258  &  48.12  &  --  & --  &  1.43$\pm$0.23  & 3.72$\pm$0.44\\
43.31+59.2\tablefootmark{a}\tablefootmark{b}\tablefootmark{c}\tablefootmark{d}  &  52.020  &  45.40  &  --  & --  &  4.59$\pm$0.49  & 5.01$\pm$0.57\\
45.42+67.4\tablefootmark{a}\tablefootmark{b}\tablefootmark{c}\tablefootmark{d}  &  54.126  &  53.54  &  10.88$\pm$1.15  & 17.02$\pm$1.91  &  11.91$\pm$1.21  & 15.98$\pm$1.63\\
45.74+65.2\tablefootmark{a}\tablefootmark{b}\tablefootmark{c}\tablefootmark{d}  &  54.453  &  51.38  &  5.13$\pm$0.65  & 8.14$\pm$1.01  &  5.42$\pm$0.58  & 9.05$\pm$0.96\\
46.52+63.9\tablefootmark{a}\tablefootmark{b}\tablefootmark{c}\tablefootmark{d}  &  55.208  &  49.99  &  1.70$\pm$0.37  & 1.88$\pm$0.60  &  2.33$\pm$0.30  & 3.08$\pm$0.43\\
46.56+73.8\tablefootmark{a}\tablefootmark{b}\tablefootmark{c}  &  55.254  &  59.87  &  5.75$\pm$0.68  & 7.97$\pm$0.98  &  6.76$\pm$0.70  & 8.61$\pm$0.91\\
Uncatalogued  &  55.824  &  55.68  &  --  & --  &  1.05$\pm$0.20  & 1.26$\pm$0.31\\
Uncatalogued  &  57.017  &  58.75  &  --  & --  &  1.30$\pm$0.20  & 1.78$\pm$0.30\\

            \noalign{\smallskip}
            \hline
         \end{tabular}
         \tablefoot{ 
	     Sources with names (given traditionally as B1950-offset) were
detected previously by \tablefoottext{a} {\cite{wills1997},}
\tablefoottext{b}{\cite{fenech2008},} \tablefoottext{c}{\cite{fenech2010}} or
\tablefoottext{d}{\cite{gendre2013}.}
	     Objects not listed in these studies were marked as
``Uncatalogued''. All uncatalogued objects except (46.42, 64.9) were clearly
detected also by eMERLIN at 1.6\,GHz. See Sect. \ref{sect:new} for further
notes on source detection.
}
   \end{table*}

\begin{figure*}[htbp]
\centering
    \includegraphics[width=0.76\textwidth]{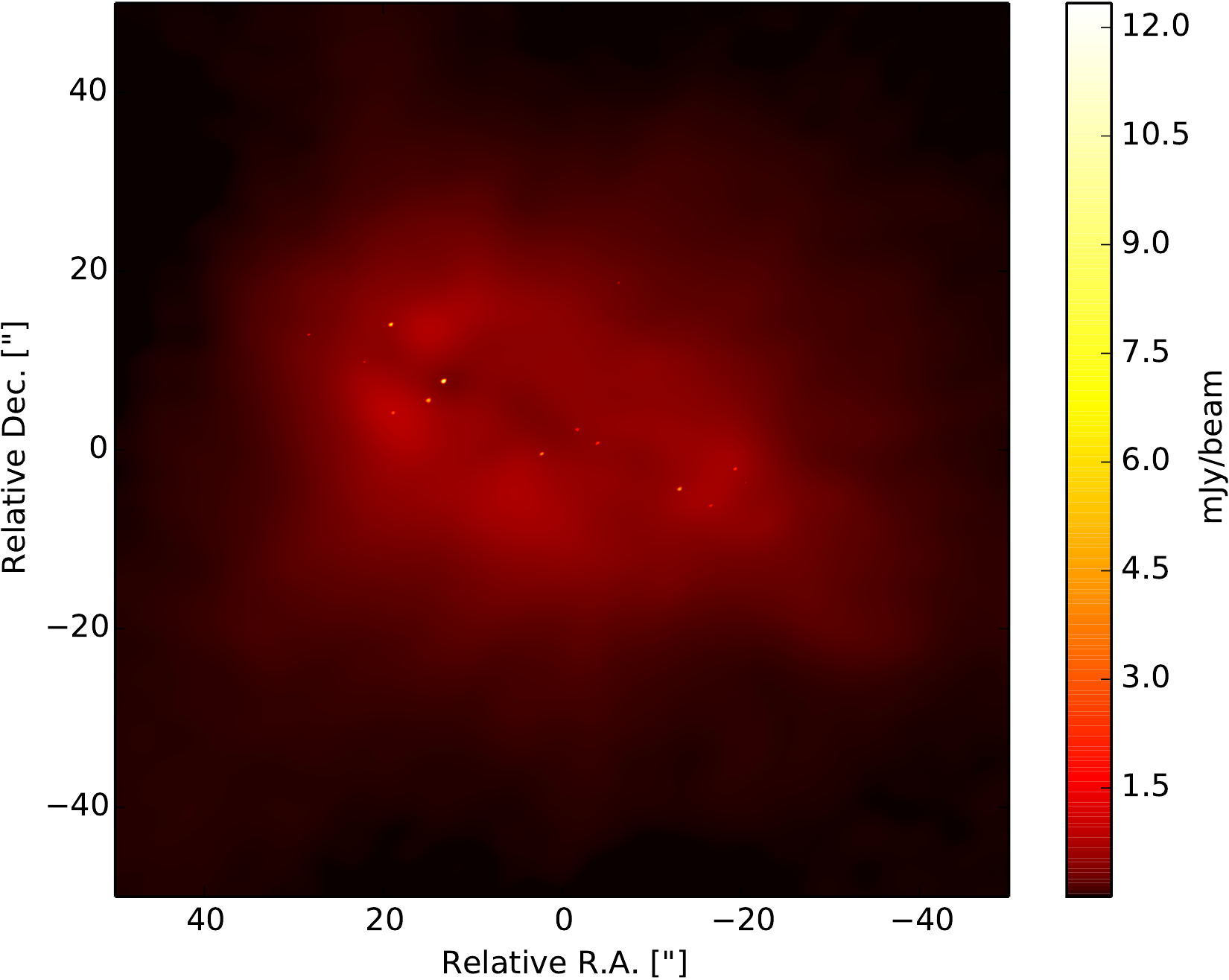}
\caption{
Combined image illustrating the relative brightness between the compact and extended
emission at 154\,MHz. This image was made by combining (using the task
\emph{feather} in CASA) the RS image (Fig. \ref{fig:M82RS}) with a thresholded
 version of the INT image, Fig. \ref{fig:M82INT}, only including emission brighter 
than $4\sigma$.  The threshold was needed to remove the relatively
strong noise in the INT image which would otherwise distort the weaker parts
of the extended emission.
Since the brightness difference between compact and extended emission is large,
a square-root colour scale is used to better show the large scale emission.  The
colour scale is in mJy/beam, where the beam is the INT beam in
Fig. \ref{fig:M82INT}, i.e., $0.36''\times0.23''$.  Positions are given relative
to the M82 position in Table \ref{tab:targetlist}.
    \label{fig:feather}
}
\end{figure*}

\subsection{The shape and flux of the diffuse emission}
The diffuse emission, Fig. \ref{fig:M82RS_BOTH},  shows a ``ring''-like
structure with a ``hole'' in the centre. The ``ring'' we interpret as coming
from emission above and below the starburst disk at the base of the large scale
outflow, perhaps more clearly visible in the combined INT+RS colour image,
see Fig.  \ref{fig:feather}.  The minimum brightness in the hole is less than
half the peak brightness of the surrounding ring.  In addition to the central
hole, we see a decrease in extended emission tracing the region where most of
the compact objects listed in Table \ref{tab:props} are located.  We interpret
this decrease as due to free-free absorption by the central star forming disk,
seen almost edge-on.  

Integrated flux densities were obtained by summing pixels above 2$\sigma$
clearly associated with M82 in Fig. \ref{fig:M82RS_BOTH}.  At 118\,MHz we obtain
12.0$\pm1.2$\,Jy with a peak of 225\,mJy/beam, and at 154\,MHz we obtain
13.5$\pm1.4$\,Jy with peak of 173\,mJy/beam. The middle hole has a minimum
brightness of 91\,mJy/beam at 118\,MHz and 70\,mJy/beam at 154\,MHz.  Although
absorption has a clear impact on some parts of the core, the integrated flux
densities measured at 118 and 154\,MHz show that the central parts of M82
are brighter than the $\sim5$\,Jy expected based on detailed modelling of M82
\citep[Fig. 5]{lacki2013}.

When imaging Fig. \ref{fig:M82RS_BOTH} we excluded baselines shorter than
2k$\lambda$, hence we miss flux present at the largest spatial scales.
It follows that the flux densities recovered here for the extended emission are to
be regarded as lower limits at the respective frequencies.  Images of M82
also including shorter baselines will be published in a forthcoming paper.

\subsection{The ``uncatalogued'' compact objects}
\label{sect:new}
In Table \ref{tab:props} we label some compact objects, seen in Fig.
\ref{fig:M82INT_BOTH}, as ``uncatalogued''.
Most sources seen with LOFAR are almost certainly old SNRs, some of which
have either not been detected previously due to insufficient sensitivity,
or have not been reported because of selection effects.
Previous studies have used different selection criteria for when to
report a detection. For example, \cite{wills1997} reported sources detected at
408\,MHz only if they were brighter than 1\,mJy at 5\,GHz.  In fact, a
preliminary re-analysis of the original Wills 408\,MHz data show evidence of
emission from three objects reported as ``uncatalogued'' in Table
\ref{tab:props}: (50.820, 46.60; 2.1\,mJy), (55.824, 55.68; 1.2\,mJy), and
(57.017, 58.75; 1.8\,mJy).
We note, by visual inspection of the recent 1.6\,GHz eMERLIN observations, that
all but one (at position 46.42, 64.9 in Table. \ref{tab:props}) of the uncatalogued objects detected with LOFAR at 154\,MHz are
also present at 1.6\,GHz.  This strongly suggests that these objects
are real. 

\subsection{Do we detect H{\sc II} regions?}
\label{sect:HIIlimit}
To calculate a lower limit on the brightness temperatures of the detected
sources we can use eq. 5 by
\cite{condon1991}: $T_b=(c^2S_I/2k\nu^2)\cdot(8\text{ln}(2)/3\pi\theta_M
\theta_m)$. Assuming a uniform elliptical source component with angular diameters
$\theta_M=0.36'',\theta_m=0.23''$ (beam size), and integrated flux density
5$\times$0.15\,mJy (detection threshold), we find a lower limit of $10^{5.5}$\,K
for the compact sources detected at 154\,MHz.  Thermal (free-free) emission from ionised
gas in H{\sc II} regions have brightness temperatures up to $2\cdot10^4\mathrm{\,K}$ 
\citep[Sect.\,11.2.1]{tools}.  The high resolution of our image thus
allows us to discard the hypothesis that the objects listed in Table
\ref{tab:props} are simple H{\sc II} regions.

\subsection{Low-frequency spectra of compact objects}
\label{sect:lowfreq}
Our observations provide important constraints on the (likely) foreground
free-free absorption due to ionised gas, not only of the detected objects in
Table \ref{tab:props}, but also as upper limits on all sources seen at higher
frequencies but not detected here.  To illustrate this we chose the five
brightest objects at 1.7\,GHz listed by \cite{fenech2010}: 41.95+57.5
(38.25\,mJy), 43.31+59.2 (23.55\,mJy), 45.17+61.2 (17.60\,mJy) 44.01+59.6
(12.32\,mJy), and 40.68+55.1 (11.33\,mJy). Of these, only 43.31+59.2 is 
detected at 154\,MHz (5\,mJy). 
We note in particular that the brightest object at 1.7\,GHz, 41.95+57.5, was
measured to be 100\,mJy at 408\,MHz by \cite{wills1997} but we do not detect any emission
with LOFAR at this position.
The spectrum of this source is, however, known to evolve over time and significant changes 
are expected during the two decades since the 408\,MHz data were taken. Based on 4.8\,GHz observations, \cite{allen1998} 
derive a decay rate of 8.8\%yr$^{-1}$ for this object, but the decay estimate is uncertain for lower frequencies.
Future modelling of this object may use these LOFAR measurements to better constrain the spectral evolution
also at low frequencies.

\subsubsection{Sources detected also at 408\,MHz}
\begin{figure*}[t]
\centering
    \includegraphics[width=\textwidth]{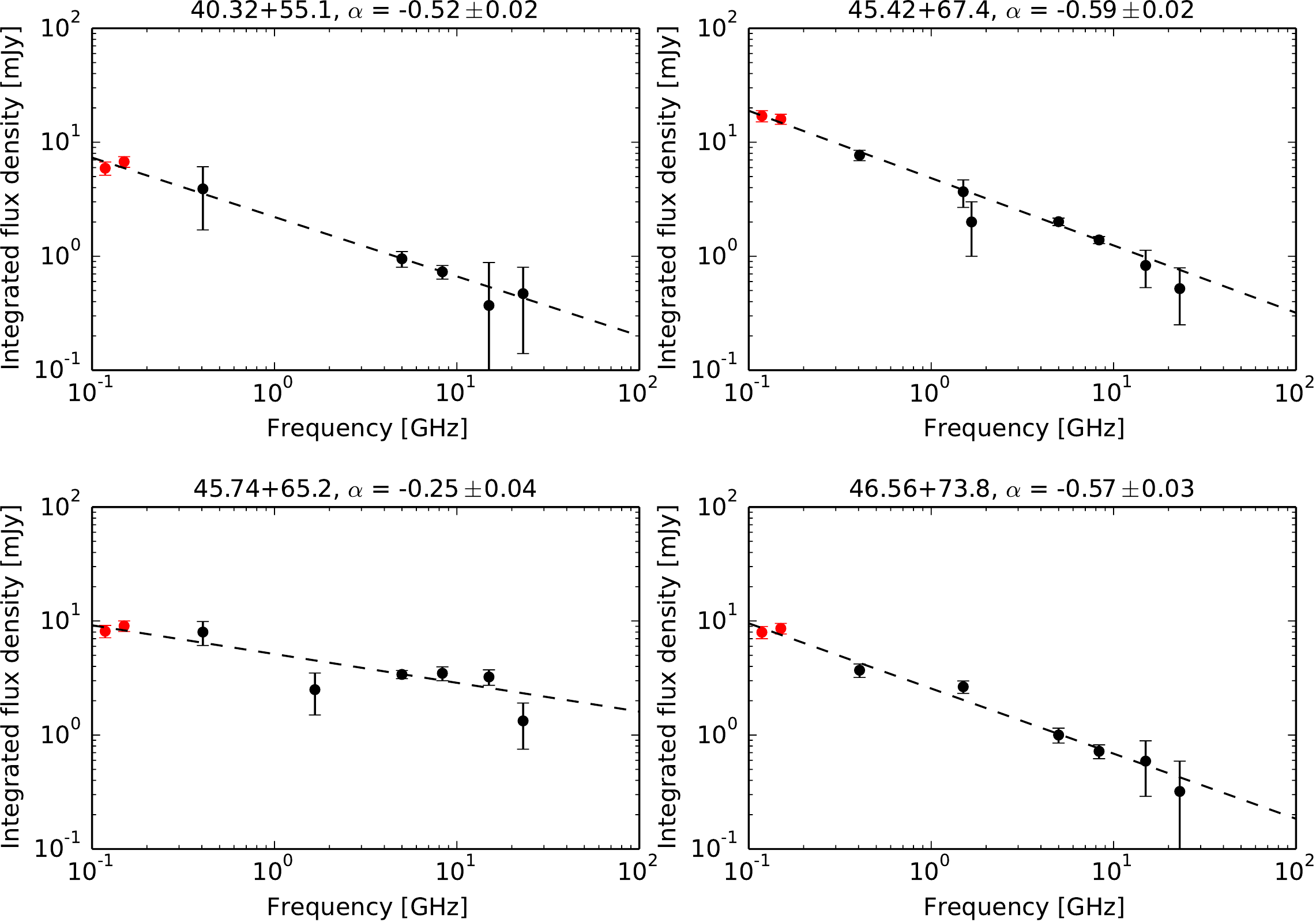}
\caption{
    Spectra of four compact objects detected with LOFAR, and also reported at 408\,MHz by \cite{wills1997}. These objects show no low-frequency turn-over
at LOFAR frequencies.  LOFAR data points are plotted in red. Non-LOFAR data points are VLA and MERLIN, as presented
in Table 2 by \cite{allen1998}.  A simple power law
($S_\nu\propto\nu^\alpha$) was fitted and is plotted as a dashed line. The best
fit $\alpha$ is given in the respective figure title.
}
    \label{fig:powerlawspectra}
\end{figure*}

To connect these results to the previous subarcsecond images of M82 at
408\,MHz, we selected the nine sources listed in Table \ref{tab:props} where
408\,MHz flux densities were reported by \cite{wills1997}.  We find that the
spectra of four objects can be fitted by a power law (see Fig.
\ref{fig:powerlawspectra}) i.e., no evidence of spectral turnovers down to
118\,MHz.  Three of these sources have spectral indices typical of supernova
remnants or young supernovae ($\alpha\approx-0.5$) but one object (45.74+65.2)
has a flatter spectrum ($\alpha\approx-0.3$); these objects are discussed in 
Sect. \ref{sect:flat}.

\begin{figure*}[t]
\centering
    \includegraphics[width=\textwidth]{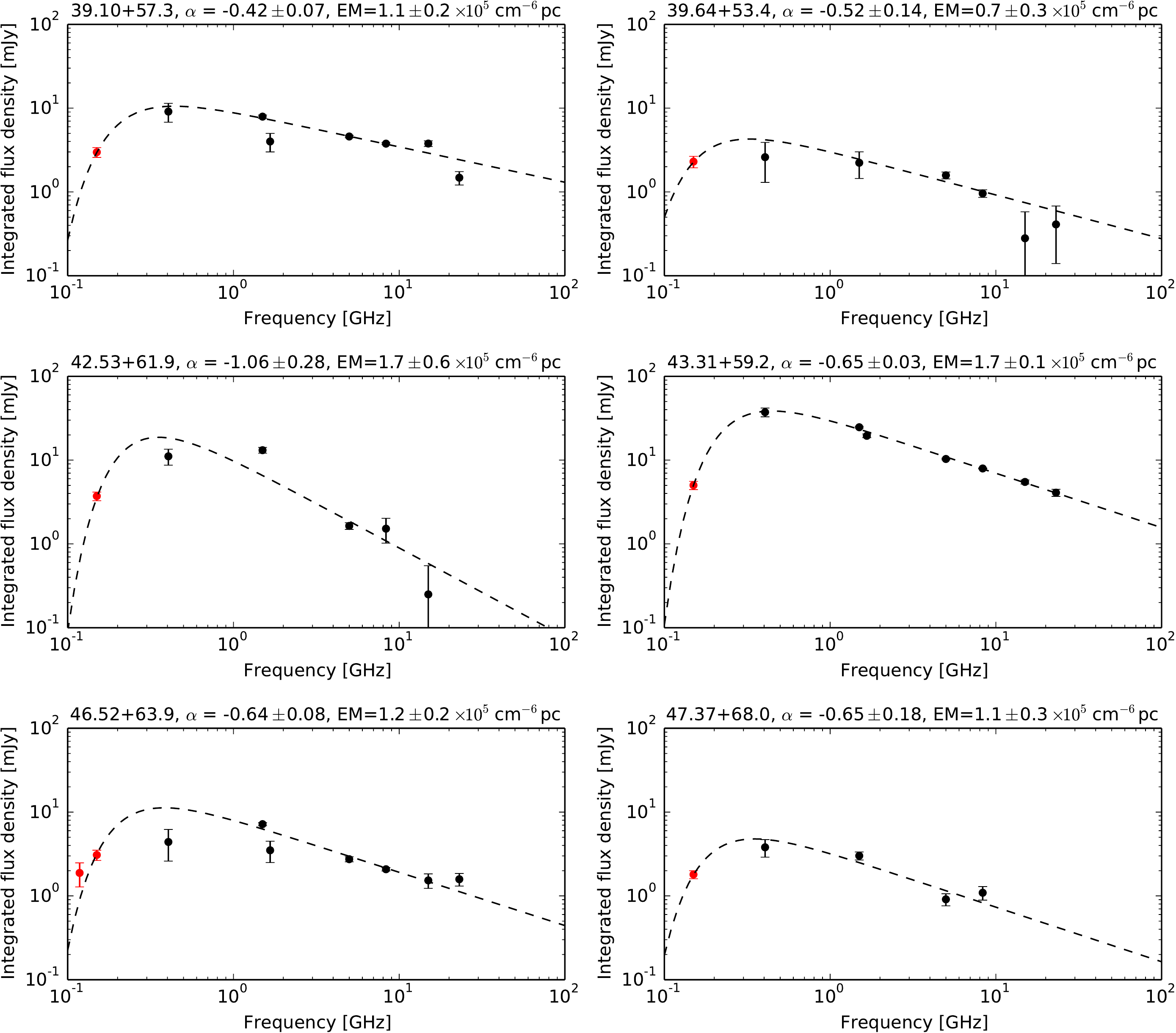}
\caption{
    Spectra of six objects detected with LOFAR, and also reported 
at 408\,MHz by \cite{wills1997}. Five of these are compact and listed in Table \ref{tab:props}, while
``47.37+68.0'' is resolved and discussed in Sect. \ref{sect:weakbutextended}.
These objects show evidence of a low-frequency turn-over
at LOFAR frequencies.  LOFAR data points are plotted in red.  Non-LOFAR data points are VLA and MERLIN, as presented
in Table 2 by \cite{allen1998}.  
A model of SNR spectra was fitted to each object as described in
Sect. \ref{sect:lowfreq}.  The best fit spectral index $\alpha$ and emission measure
$\mathrm{EM}$ are given in the respective figure title.  }
\label{fig:turnoverspectra}

\end{figure*}

The remaining five sources show indications of a low-frequency turnover, see
Fig.  \ref{fig:turnoverspectra}.  
We have attempted to fit the spectra of these
sources using the model used by \cite{wills1997}.  In this model, the
emission comes from a background source (i.e. the SNR) with spectral index $\alpha$ and is
attenuated by foreground free-free absorption.  The spectrum
$S_\nu$ is described as $S_\nu\propto\nu^\alpha e^{-\tau}$, where
$\tau=8.2\times10^{-2}\nu^{-2.1}\mathrm{EM}/T_e^{1.35}$ and $\mathrm{EM}$ is
the emission measure in cm$^{-6}$pc, $\nu$ the frequency in GHz and $T_e$ is
the electron temperature, taken to be 10\,000\,K. 
A more detailed analysis based on fitting also upper limits for many more
sources will be presented in a forthcoming paper.

The best fitting model was found using standard least-squared fitting in SciPy
and the best fit parameters $\alpha$ and EM are given in the respective figure
title.  For these objects LOFAR provides a very important addition to the
spectrum and makes it possible to determine the emission measure with higher
precision than before. We note that the fitted EM values are in the relatively
narrow range 0.7-1.7$\cdot10^{5}$\,cm$^{-6}$pc, but given the small number
statistics we refrain from further analysis of the EM range. 
We note that detailed models of SNR evolution may differentiate between internal 
and external free-free absorption mechanisms, see for example \cite{delaney2014} 
for Cassiopeia A. Unfortunately we do not yet have enough measurements 
of the sources in M82 to properly separate these effects.

With bright SNRs it is also possible to study the cold, diffuse ISM through absorption 
as radio recombination lines (RRLs). This was recently done in M82 using LOFAR LBA by 
\cite{morabito2014} although this detection only used the shortest LOFAR baselines and 
hence could not spatially resolve and locate the RRLs within M82. 
In a future publication we aim to map the RRLs detected by \cite{morabito2014} 
by using the techniques explained in this paper for calibration and imaging of international 
and remote baselines, as well as search for and map RRLs within the HBA data presented here. 

\subsubsection{Flattening of spectra}
\label{sect:flat}
In the objects 42.35+61.9 and 46.52+63.9 the model does not represent the data
very well.  The measured spectra of these objects are flatter and the turn-over
frequency higher than the best-fit model.  
A flattening of the spectrum (as also seen in 45.74+65.2) can be explained by
ejecta-opacity effects in the shell of a radio supernova remnant, as observed
in SN1993J by \cite{martividal2011SN1993J}.  In this case, the ejecta becomes
optically thin at higher frequencies and allow us to see emission also from the
rear side of the shell at higher frequencies, hence flattening the spectrum.
The source with the flattest spectrum, 45.74+65.2, is clearly a supernova
remnant as seen in \cite{fenech2010}, their Fig. 3, and since it is bright
(third brightest at 154\,MHz) ejecta-opacity effects could be important.
Another possible explanation is that this object is a plerion-shell composite,
see e.g. \cite{weiler1988}. More advanced modelling of these spectra will be
presented in a future paper.

\subsubsection{The eMERLIN transient 43.78+59.3}
A ``transient'' source was reported by \cite{muxlow2010} at position
$09^h55^m52^s.5083$ and 69$^\circ40'45''.410$ (J2000).  
At 118\,MHz we do not detect anything at this position. At 154\,MHz we find a 
weak (0.45\,mJy/beam) signal at position $09^h55^m52^s.520$, 
69$^\circ40'45''.43$ (J2000). This may be associated with 43.78+59.3 but
since it is so weak (only 3$\sigma$) we do not claim a detection and treat 
this as an upper limit.

\subsubsection{No detection of SN2008iz} SN2008iz \citep{brunthaler2009} is not
detected in these observations; the maximum value which could be associated
with this source is 0.90\,mJy/beam at 118\,MHz and 0.41\,mJy/beam at 154\,MHz,
i.e. $<3\sigma$.  This source was originally predicted to be $140$\,mJy at
154\,MHz (at the time of this observation), using the standard mini-shell model
for radio SNe \citep{chevalier1982,chevalier1982-2,weiler2002}.  However, recent elaborate
multi-frequency modelling using all available data for this object, and adding
a low-frequency cutoff due to the Razin-Tsytovich effect (the suppression of
the emission below a certain frequency due to plasma propagation effects),
predict a flux density at 154\,MHz of $10^{-4}$\,mJy at the time of these
observations (Martí-Vidal et. al 2014, in prep.), i.e well below our detection
limit.  

\subsection{Four weak resolved objects} 
\label{sect:weakbutextended} In addition to the compact objects listed in Table
\ref{tab:props} we find four resolved features in Fig. \ref{fig:M82INT} which,
although weaker than our point source sensitivity of $5\sigma$, are above
several sigma over multiple beams and/or correspond to structure seen at higher
frequencies and which we are confident are real.  Cutouts of each of these
features are shown in Figs.  \ref{fig:leftshell} to \ref{fig:extended}.

\subsubsection{Two weak shell-like features}
Two weak shell-like features, with diameters $\sim0.4''$, are seen at 154\,MHz,
see top panels in Fig. \ref{fig:zoom}. 

One (top-left panel) has central position at $09^h55^m56^s.030$,
69$^\circ40'53''.80$ (J2000) with peak flux density peak flux density
0.6\,mJy/beam and integrated flux density 1.8\,mJy/beam (summing by eye pixels
associated with the source).  We identify this object as the source ``47.37+
68.0'' as listed by \cite{wills1997}. This object was not included in Table
\ref{tab:props} since the peak is formally below our detection threshold.
However, given the clear shell-like structure, the significant integrated flux
density, and the good match with previous observations, we include
this source in the spectral fits presented in Fig. \ref{fig:turnoverspectra},
with a flux density of 1.8$\pm$0.2\,mJy at 154\,MHz. The fitted spectrum
indicates a low-frequency turnover below 408\,MHz, see bottom-right panel in
Fig.  \ref{fig:turnoverspectra}).

The second shell-like feature (top-right panel) has centre position at
$09^h55^m53^s.100$, 69$^\circ40'49''.10$ (J2000) with peak brightness of 
0.7\,mJy/beam and integrated flux density 1.8\,mJy at 154\,MHz, and a peak
of 0.1\,mJy/beam and integrated flux density of 0.3\,mJy at 1.6\,GHz.  It is
hard to accurately measure the integrated flux density at 1.6\,GHz for this
object because of nearby large-scale emission sampled with eMERLIN at this
frequency.  This object is very similar in structure and flux density at
154\,MHz to ``47.37+68.0'', but we could not find a corresponding source
catalogued in either of \cite{wills1997}, \cite{fenech2008}, \cite{fenech2010}, or \cite{gendre2013}.
This object is probably an SNR which, because of the nearby confusing
large-scale emission, has not previously been clearly identified. 

\subsubsection{A peculiar linear feature}
\label{sect:linear}
\begin{figure*}[htbp]
\centering
\subfigure[SNR ``47.37+68.0''. Contour at $4\sigma_{1.6}$.]{
    \includegraphics[width=0.45\textwidth]{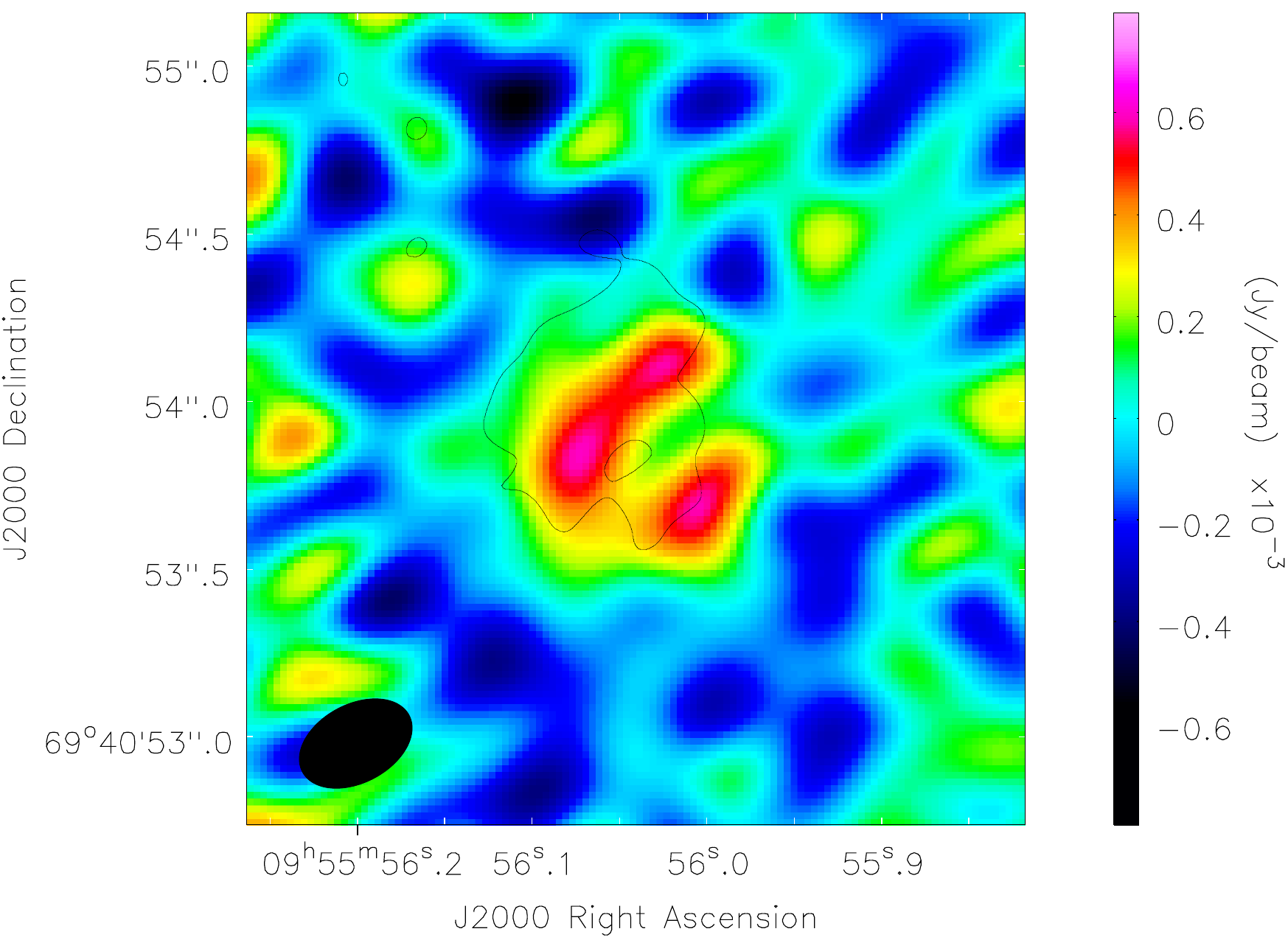}
    \label{fig:leftshell}
}
\subfigure[A shell-like feature, probably SNR. Contour at $4\sigma_{1.6}$.]{
    \includegraphics[width=0.45\textwidth]{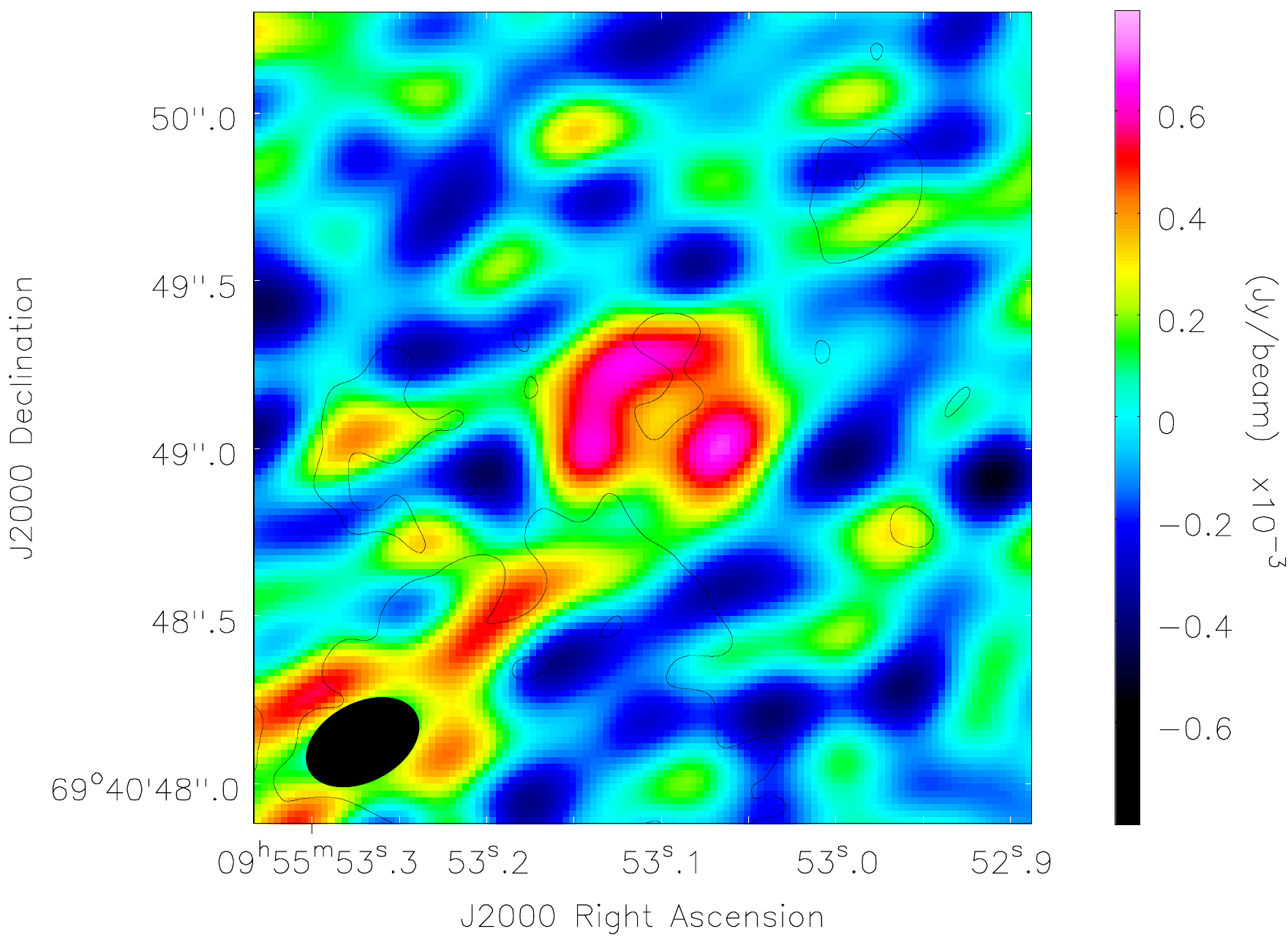}
    \label{fig:rightshell}
}
\subfigure[A linear feature. Contour at $2.5\sigma_{1.6}$.]{
    \includegraphics[width=0.45\textwidth]{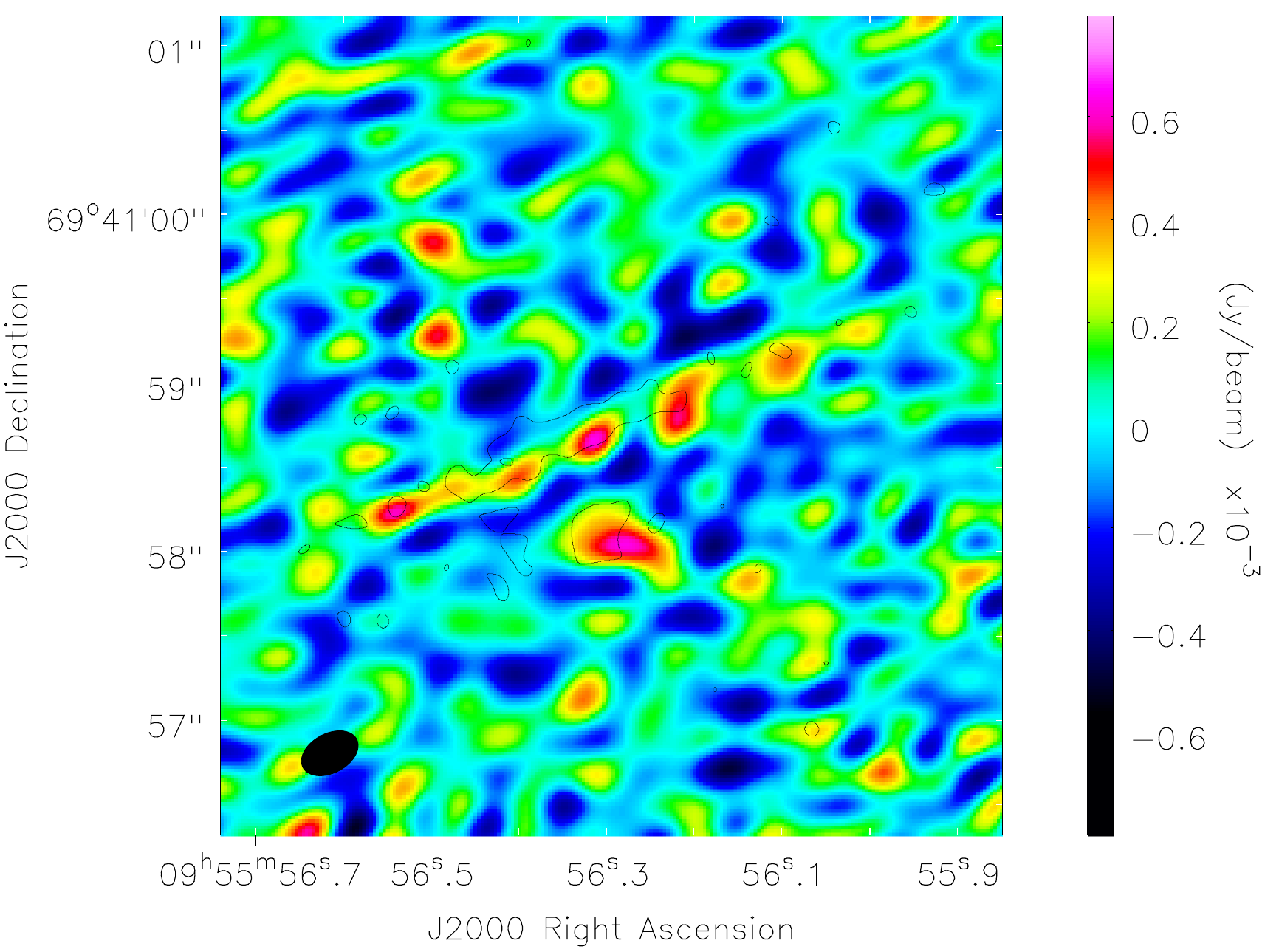}
    \label{fig:linear}
}
\subfigure[A resolved but not clearly shell-like feature. Contour at $4\sigma_{1.6}$.]{
    \includegraphics[width=0.45\textwidth]{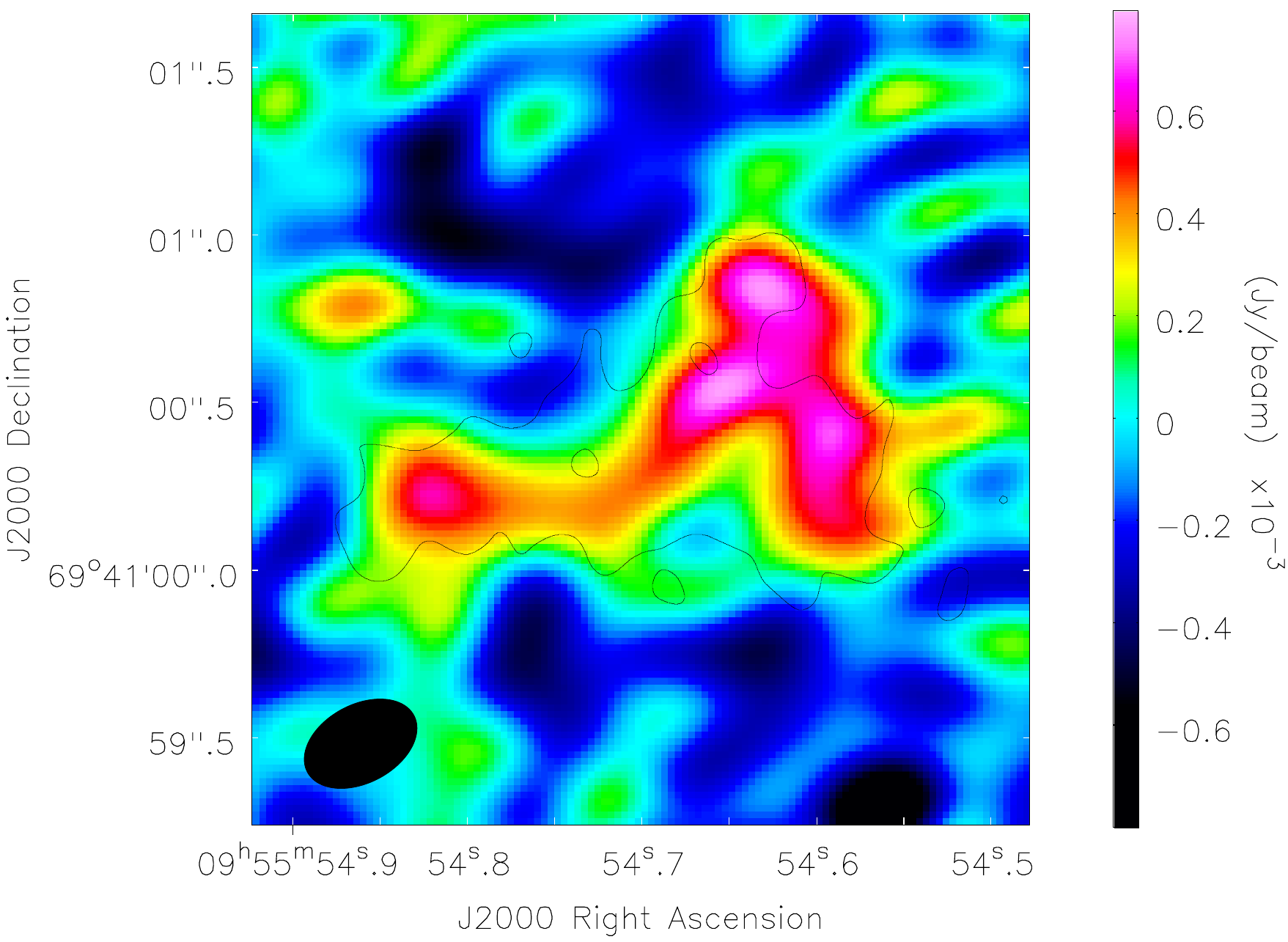}
    \label{fig:extended}
}
\caption{ Zoom on four weak but resolved features seen in Fig.
\ref{fig:M82INT}. The LOFAR 154\,MHz-beam is shown in the lower left corner.
Each feature is briefly discussed in Sect.  \ref{sect:weakbutextended} where
flux densities and positions are given.  Colours are LOFAR 154\,MHz, contours
are eMERLIN at 1.6\,GHz ($\sigma_{1.6}=16\,\mu\mathrm{Jy/beam}$,
\citealt{miguel2014}), see respective panel text for contour levels.  The top
panels show two shell-like structures; the top-left is identified as the SNR
listed as ``47.37+68.0'' by \cite{wills1997}, while the top-right is
uncatalogued.  The bottom-left panel shows a linear feature and bottom-right
shows a resolved but not clearly shell-like object. Note that the bottom-left
panel covers a larger area on sky, and has lower contour levels, compared to
the other three panels.  We note a minor positional offset of approximately
60\,mas between the LOFAR and eMERLIN images, see Sect. \ref{sect:posacc} for a
brief discussion on positional accuracy. \label{fig:zoom}
}
\end{figure*}

In the north-east part of the Fig. \ref{fig:M82INT} we see a linear feature
(zoom in Fig. \ref{fig:linear}) extending $3''$ (50\,pc) i.e. for 10 beams, not
aligned with the disk of M82.
This object has an integrated flux density of 2.1\,mJy and a peak of
0.6\,mJy/beam (only seen at 154\,MHz).  Although formally below our detection
limit, this object was also detected in recent eMERLIN observations at
1.6\,GHz, see Fig.  \ref{fig:linear}, with an integrated flux density of
1.5\,mJy.  This feature is similar in size (50\,pc) to the non-thermal radio
filaments seen prominently in the direction perpendicular to the Galactic
plane\footnote{See also http://apod.nasa.gov/apod/ap080427.html.}
\citep{yusef1987}.
However, these radio filaments were observed by \cite{law2010} to have flux
densities of $\sim20-200$\,mJy at 20\,cm wavelength which, given the distance
ratio of $\sim 400$ would appear as $\sim 1$\,$\mu$Jy in M82.
The observed flux density at 1.6\,GHz is more than a thousand times the
luminosity expected if this was a non-thermal filament similar to those in the
centre of the Milky way.  We refrain from further speculation about the nature
of this source.

\subsubsection{A resolved but not clearly shell-like object}
\label{sect:extsource}
At position $09^h55^m54^s.62$ and $69^\circ41'00''.5$ (J2000) is a clearly
resolved object (see zoom in Fig.  \ref{fig:extended}) with peak flux density
0.8\,mJy/beam and integrated flux density 4.5\,mJy at 154\,MHz.  With chosen
input parameters, PyBDSM did not detect it (parameters given in Sect.
\ref{sect:randd}), but it is clearly seen also in recent eMERLIN 1.6\,GHz
observations where an integrated flux density of 5.5\,mJy and  a peak
brightness of 140\,$\mu$Jy/beam (eMERLIN beam $150''\times126''$) is recovered
for this source.  This object was not seen by either of
\cite{wills1997,fenech2008,fenech2010,gendre2013}.  However, as discussed in
Sect. \ref{sect:new}, this resolved object is probably old but since it is
faint it might not have been detectable without the now improved sensitivity of
eMERLIN.

\subsection{Implications for LOFAR  International Baseline Imaging} 
The results presented in this paper show that  sensitive ($\sigma=
0.15\mathrm{\,mJy/beam}$) random noise-limited images of complex
sources can be made at HBA frequencies with international LOFAR baselines. The
images obtained set a new record in terms of combined resolution and sensitivity 
for science images at frequencies below 327\,MHz. 
These data were reduced in a relatively straightforward manner
using conventional VLBI imaging techniques.  LOFAR international baseline
imaging, although often perceived as more difficult than imaging on Dutch
baselines, can in fact be less computationally expensive and algorithmically
complex. 
International baseline imaging is in many aspects simpler than Dutch baseline imaging, 
because we can generally ignore interference from other bright
sources in the sky.  The effect of one source on the position of another is
greatly reduced by time and frequency coherence losses such that each compact
source can be imaged independently in small fields.

For a distant potentially interfering source, the interfering contribution to
the target source visibility declines by factors of $u^{-1}$ as a function of
baseline length ($u$)  for both frequency and time decorrelation.  Furthermore,
for  most  of the source population resolved on longer Dutch LOFAR baselines,
the intrinsic visibility structure decreases faster than $u^{-2}$ (a rough 
approximation based on experience of typical dependence of visibility amplitude 
vs baseline length for resolved radio sources), giving a total decrease of
interfering signals with baseline length scaling as $ u^{-4}$. This means  that
the effect of interfering signals is a million times less for baselines of
1000\,km as opposed to 30\,km. This effect explains why the influence of the
brightest sources at LOFAR frequencies, like Cassiopeia A or Cygnus A, can be ignored
at international baseline resolution, as can most Jansky level sources within
the station beam.  

The small field of view regime is where cm-VLBI usually operates and this
regime greatly simplifies imaging. In this regime, target images are generally
smaller than the isoplanatic patch, and therefore only a single-direction
station-dependent correction needs to be determined.  Likewise, over such small
fields, $w$-term effects and station beam variations can generally be ignored.
This is different from LOFAR core-resolution imaging, where in order to get
noise-limited (rather than ``dynamic range''-limited) images at any given point
in a field the whole field must be imaged using multi-directional calibration
techniques.  In the intermediate regime of LOFAR Dutch remote ($<100$\,km)
baselines we have shown that on a very bright target source such as M82 it is
also possible to get useful science images by standard VLBI techniques, as
demonstrated by Fig.  \ref{fig:M82RS}.  Further increasing the image sensitivity by
deconvolving interfering sources may require other techniques.

Whether one can produce useful images using the small-field approximation and
VLBI software depends on the brightness of the target source and, if necessary,
the availability of close sources to use as calibrators similar to what is done
in cm-VLBI.  A new pipeline is being developed by \cite{moldon2014} to help
quickly identify nearer calibrators, which should make the calibration process
more robust by reducing the degree of spatial extrapolation.

\subsection{Future data analysis and observations}
This paper has concentrated on presenting the imaging results applying VLBI
techniques to the analysis of international and Dutch remote baselines. We have
shown that these techniques can be used to obtain high dynamic range images
close to the thermal noise.  Based on Fig. \ref{fig:M82RS} the positions of
SNRs showing low-frequency turnovers and SNRs not showing a turnover are
distributed randomly with respect to stronger and weaker absorption features in
the diffuse emission.  This suggests that the free-free absorbing ionised
medium is clumpy rather than uniform, as discussed by \cite{lacki2013}.  
However, this can be better constrained by also modelling the spectrum of the
diffuse emission.
A subsequent analysis paper will jointly
model the spectra of the compact sources and the diffuse emission in which they
are embedded.  Other data sets on M82 have been observed, optimised to image
the extended halo structure at HBA and LBA frequencies and will be presented
elsewhere (Adebahr et al, in prep.).  This paper only presents results on M82,
the M81* data is yet to be fully analysed and we will attempt in future work to
characterise the low-frequency structure of the M81 nucleus, detect or set
limits on the flux density of SN1993J, and try to detect other supernova
remnants in M81.

It is clear from the observations presented here that there are many compact
objects in M82 at, or just below, our sensitivity detection level.  Modest
improvements in sensitivity, achievable in follow-up observations, will allow
us to detect more sources. It should be noted that the observational setup used
for the observations presented in this paper was designed in order to minimise
the data processing needed rather than to maximise sensitivity.  In particular,
the bandwidth was split equally three ways between M82, M81* and the bright
calibrator, and each was allocated a separate station beam. The sources M81*
and M82 are close enough that they are well within the same station beam, and
can in subsequent observations be observed simultaneously within one beam at
full bandwidth.  In addition to this, we can use the procedure suggested by
\cite{moldon2014} to identify a closer primary calibrator for M82, which can also
be observed in the same beam.  To use this ``three source in one beam''
-technique we need to store initially the data after correlation with high
temporal and spectral resolution, phase rotate the data to the position of the
two sources, and average in time and frequency. This is conceptually simple but
requires processing large volumes of (temporary) data. Efforts are underway to
implement and automate such a pipeline \citep{moldon2014}.  As a
practical matter such processing should therefore be done immediately after
correlation on the correlator output cluster, before shipping data to other
facilities.

The results in this paper show that the prospects for science exploitation of LOFAR
international baseline data  are excellent and the data analysis in principle
straightforward. To make it more exploitable by the general community,
software for non-experts is being implemented. A significant issue
that has to date limited the exploitation of LOFAR international baselines has
been the reliability of the data links between the international stations and
the correlator (the international stations themselves show high reliability)
but recently much progress has been made in solving this issue.
For optimal sensitivity and speed in the calibration process it is also desirable
to use more general full-bandwidth fringe-fitting strategies that fully include 
non-dispersive and dispersive delays as well as differential Faraday rotation.
Software to enable such calibration techniques is currently under development.

\section{Summary}
\label{sect:conclusions}
The International LOFAR telescope has been used to make subarcsecond
resolution images of the sky at 118\,MHz and 154\,MHz.  Our 154\,MHz continuum
image using international baselines, with a synthesised beam of
$0.36''\times0.23''$ and RMS noise 0.15\,mJy/beam, is a new record in terms of
combined image resolution and point-source sensitivity for science images at
frequencies below 327\,MHz.  

In our high-resolution image, we detect 16
compact objects above 5$\sigma$ at 154\,MHz within the inner $1'$ of M82. Six of
these are also detected above 5$\sigma$ at 118\,MHz.
Four resolved features are seen (but with peak flux densities below $5\sigma$) 
in the high-resolution image. These are also seen with eMERLIN at 1.6\,GHz.

In contrast to predictions of the standard RSNe model
\citep{chevalier1982,chevalier1982-2,weiler2002}, we do not detect any emission
($<3\sigma$) from supernova 2008iz.  However, new modelling, accounting for
significant plasma effects in the shocked region, is in good agreement with our
observations.
We do not detect ($<3\sigma$) the eMERLIN transient source 43.78+59.3, nor
the object 41.95+57.5 which is the brightest compact object at 1.7\,GHz.
Using data on Dutch (remote) baselines, we find diffuse emission
surrounding compact objects with total integrated flux density 12.0$\pm1.2$\,Jy
at 118\,MHz and 13.5$\pm$1.4\,Jy at 154\,MHz. Significant absorption is seen
along the star forming disk where most of the compact objects are located. The
strongest absorption is seen towards the centre of M82 as a ``hole'' in the
diffuse emission.

Further analysis, also using upper limits imposed by these observations
on sources not detected here, will make it possible to trace the spatial
structure (uniform/clumpy) of the absorbing ionised gas component present in
the nucleus of M82.

\begin{acknowledgements}
We acknowledge assistance by the LOFAR science support as well as Simon Casey
for installing LOFAR software in Onsala. 
This research has made use of the NASA/IPAC Extragalactic Database (NED) which
is operated by the Jet Propulsion Laboratory, California Institute of
Technology, under contract with the National Aeronautics and Space
Administration.  The research leading to these results has received funding
from the European Commission Seventh Framework Programme (FP/2007-2013) under
grant agreement No 283393 (RadioNet3).
e-MERLIN is the UK's National Radio Interferometric facility, operated by the
University of Manchester on behalf of the Science and Technology Facilities
Council (STFC).
LOFAR, the Low Frequency Array, designed and constructed by ASTRON, has
facilities in several countries, that are owned by various parties (each with
their own funding sources), and that are collectively operated by the
International LOFAR Telescope (ILT) foundation under a joint scientific policy.

\end{acknowledgements}

\bibliographystyle{aa}
\bibliography{paper}

\Online

\end{document}